\documentclass[fleqn,usenatbib]{mnras}

\usepackage{newtxtext,newtxmath}
\usepackage[T1]{fontenc}

\DeclareRobustCommand{\VAN}[3]{#2}
\let\VANthebibliography\thebibliography
\def\thebibliography{\DeclareRobustCommand{\VAN}[3]{##3}\VANthebibliography}

\usepackage{graphicx}	% Including figure files
\usepackage{amsmath}	% Advanced maths commands\
\usepackage{hyperref}
\usepackage{natbib}
\usepackage{caption}
\usepackage{multirow}
\usepackage{geometry}
\usepackage{booktabs}
\usepackage[table]{xcolor}
\usepackage{array}
\usepackage{mathtools}
\usepackage{xspace}

\usepackage{adjustbox}  % For the orcid ids

\newcommand{\Reff}{\ensuremath{r_{\mathrm{e}}}\xspace}

\newcommand{\vcirc}{\ensuremath{v_{\mathrm{circ}}}\xspace}

\newcommand{\vrot}{\ensuremath{v_{\mathrm{rot}}}\xspace}

\newcommand{\Msun}{\ensuremath{\rm M_{\odot}}\xspace}

\newcommand{\Mstar}{\ensuremath{M_{\star}}\xspace}
\newcommand{\Mgas}{\ensuremath{M_{\mathrm{gas}}}\xspace}
\newcommand{\Mbar}{\ensuremath{M_{\mathrm{bar}}}\xspace}
\newcommand{\Mdyn}{\ensuremath{M_{\mathrm{dyn}}}\xspace}
\newcommand{\Mbh}{\ensuremath{M_{\mathrm{BH}}}\xspace}

\newcommand{\logMstar}{\ensuremath{\log(\Mstar ~\rm [\Msun])}\xspace}
\newcommand{\logMbh}{\ensuremath{\log(\Mbh ~\rm [\Msun])}\xspace}
\newcommand{\logMdyn}{\ensuremath{\log(\Mdyn ~\rm [\Msun])}\xspace}

\newcommand{\fgas}{\ensuremath{f_{\mathrm{gas}}}\xspace}  

\newcommand{\fDM}{\ensuremath{f_{\mathrm{DM}}}\xspace}

\newcommand{\Ha}{\ensuremath{\mathrm{H}\alpha}\xspace}

\newcommand{\geko}{\textsc{geko}}

\newcommand{\sersic}{S\'{e}rsic }
\newcommand{\angstrom}{\mbox{\normalfont\AA}}

\newcommand{\deltaMS}{\ensuremath{\Delta \rm MS}\xspace} 

\newcommand{\rotsupp}{\text{v}/\sigma_0} %_{\text{re}}
\newcommand{\disp}{\sigma_0}

\newcommand{\sigmabar}{\ensuremath{\Sigma_{\rm bar}}\xspace}

\newcommand{\logsigmabar}{\ensuremath{\log (\Sigma_{\rm bar}) [\Msun \thinspace \rm kpc^{-2}]}\xspace}

\usepackage{adjustbox}  % For the orcid ids
\newcommand\orcid[1]{\href{http://orcid.org/#1}{\adjustbox{trim={-.15\width} {0\height} {-.15\width} {0\height},clip}{\includegraphics[height=10pt]{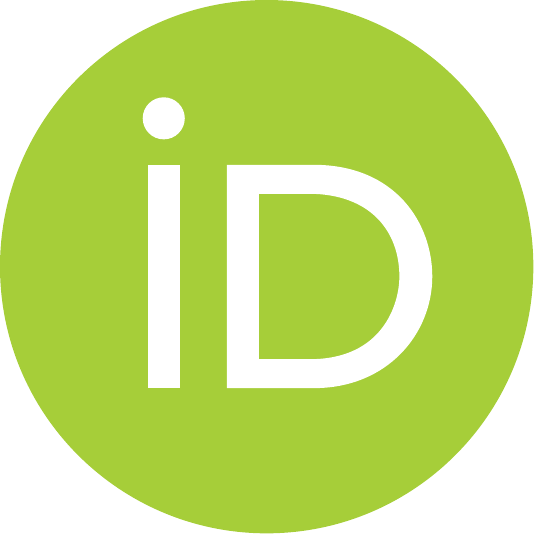}}}}

\title[DM fractions of high-z galaxies]{The dark side of early galaxies: \geko\ uncovers dark-matter fractions at 
$z\sim4-6$}

\author[A. L. Danhaive et al.]{
A. Lola Danhaive\orcid{0000-0002-9708-9958}$^{1,2}$\thanks{ald66@cam.ac.uk},
Sandro Tacchella\orcid{0000-0002-8224-4505}$^{1,2}$, Andrew J.\ Bunker\orcid{0000-0002-8651-9879}$^{3}$, Emma Curtis-Lake\orcid{0000-0002-9551-0534}$^{4}$,
\newauthor Anna de Graaff\orcid{0000-0002-2380-9801}$^{5}$, Francesco D’Eugenio\orcid{0000-0003-2388-8172}$^{1,2}$, Qiao Duan\orcid{0009-0009-8105-4564}$^{1,2}$, Eiichi Egami\orcid{0000-0003-1344-9475}$^{6}$, Daniel J.\ Eisenstein\orcid{0000-0002-2929-3121}$^{7}$,
\newauthor Benjamin D.\ Johnson\orcid{0000-0002-9280-7594}$^{7}$,  Roberto Maiolino\orcid{0000-0002-4985-3819}$^{1,2}$, William McClymont\orcid{0009-0009-5565-3790}$^{1,2}$, Marcia Rieke\orcid{0000-0002-7893-6170}$^{6}$,   
\newauthor Brant Robertson\orcid{0000-0002-4271-0364}$^{8}$, Fengwu Sun\orcid{0000-0002-4622-6617}$^{7}$, Christopher N. A. Willmer\orcid{0000-0001-9262-9997}$^{6}$, Zihao Wu\orcid{0000-0002-8876-5248}$^{7}$, Yongda Zhu\orcid{0000-0003-3307-7525}$^{6}$
\\
\\
$^{1}$Kavli Institute for Cosmology, University of Cambridge, Madingley Road, Cambridge, CB3 0HA, UK \\
$^{2}$Cavendish Laboratory, University of Cambridge, 19 JJ Thomson Avenue, Cambridge, CB3 0HE, UK\\
$^{3}$ Department of Physics, University of Oxford, Denys Wilkinson Building, Keble Road, Oxford OX1 3RH, UK \\
$^{4}$ Centre for Astrophysics Research, Department of Physics, Astronomy and Mathematics, University of Hertfordshire, Hatfield AL10 9AB, UK \\
$^{5}$ Max-Planck-Institut f\"ur Astronomie, K\"onigstuhl 17, D-69117, Heidelberg, Germany \\
$^{6}$ Steward Observatory, University of Arizona, 933 N. Cherry Avenue, Tucson, AZ 85721, USA \\
$^{7}$ Center for Astrophysics $|$ Harvard \& Smithsonian, 60 Garden St., Cambridge MA 02138 USA \\
$^{8}$ Department of Astronomy and Astrophysics, University of California, Santa Cruz, 1156 High Street, Santa Cruz, CA 95064, USA \\
}
\date{Accepted XXX. Received YYY; in original form ZZZ}

\pubyear{2015}

\begin{document}
\label{firstpage}
\pagerange{\pageref{firstpage}--\pageref{lastpage}}
\maketitle

\begin{abstract}
\textit{JWST}/NIRCam slitless spectroscopy enables dynamical mass measurements for typical star-forming galaxies only a billion years after the Big Bang. We model the H$\alpha$ morpho-kinematics of 163 galaxies at redshift $z\approx4$-6 from FRESCO and CONGRESS (with JADES imaging), using the $\texttt{geko}$ code, and infer rotational velocities and dispersions within $r_{\rm e}$. Our sample spans $\log M_{\star}\approx7$-10 and $\log M_{\rm dyn}\approx9$-11. Gas masses are estimated via scaling relations, yielding baryonic masses and dark-matter (DM) fractions $f_{\rm DM}(r<r_{\rm e})$ within the H$\alpha$ half-light radius. We find high median fractions of $\langle f_{\rm gas}\rangle=0.77$ and $\langle f_{\rm DM}\rangle=0.73$, where $f_{\rm gas}$ is measured with respect to the baryonic mass and $f_{\rm DM}$ with respect to the DM+baryonic mass. About two-thirds of systems are DM-dominated within $r_{\rm e}\sim0.5-1$ kpc. Both $f_{\rm gas}$ and $f_{\rm DM}$ decrease with stellar mass, consistent with simulations. The stellar Tully-Fisher relation shows a tentative offset to higher $v_{\rm circ}$ at fixed $M_{\star}$ and substantial intrinsic scatter, suggesting that the relation is only beginning to emerge at $z\sim5$. We measure a negative correlation between $f_{\rm DM}$ and baryonic surface density $\Sigma_{\rm bar}$, weaker but broadly consistent with trends at cosmic noon and at $z\sim0$. Qualitatively comparing with modified NFW profiles coupled to an empirical stellar-to-halo mass relation suggests that the lowest $f_{\rm DM}$ ($\lesssim0.4$) require cored inner DM profiles, while the highest fractions favour cuspier profiles, potentially reflecting adiabatic contraction. Overall, the elevated $f_{\rm gas}$ and $f_{\rm DM}$ at $z\gtrsim4$ are compatible with progenitors of baryon-dominated systems at $z\sim2$ and naturally anticipate overmassive black holes at fixed $M_{\star}$.

\end{abstract}


%%%%%%%%%%%%%%%%%%%%%%%%%%%%%%%%%%%%%%%%%%%%%%%%%%

%%%%%%%%%%%%%%%%% BODY OF PAPER %%%%%%%%%%%%%%%%%%

\section{Introduction}

The nature and physics of dark matter (DM) is to this day one of the biggest outstanding questions in astrophysics. In the widely adopted standard cosmological model of $\Lambda$CDM, DM haloes form from the collapse of overdensities in the early Universe, allowing baryons to accrete, cool, and form stars. This process is the basis of galaxy formation, and predicts the hierarchical growth of galaxies. However, the physics and interplay between the baryons and the DM remain remarkably poorly constrained. 

In fact, measuring the DM mass content of galaxies and their haloes observationally is a challenge, as often even the full baryonic mass is difficult to constrain. Many approaches have been adopted to indirectly constrain the galaxy-halo connection, such as abundance matching \citep{Marinoni:2002aa,Kravtsov:2004aa,Conroy:2006aa,Vale:2006aa,Tasitsiomi:2007aa} and clustering analyses \citep{Peacock:2000aa,Cooray:2002aa,Smith:2003aa,Paquereau:2025aa} based on N-body simulations. Many numerical simulations have measured the stellar-to-halo mass (SMHM) relation and its evolution with redshift \citep{Behroozi:2010aa, Moster:2010aa, Rodriguez-Puebla:2017aa, Behroozi:2019aa, Tacchella:2018aa}, although its overall normalisation and turnover mass are still debated and model-dependent. In part, this is due to the circularity of modelling inputs and outputs, as cosmological hydrodynamical simulations are usually calibrated to SMHM relations measured from abundance matching or derived from semi-empirical models. Placing direct observational constraints on this relation would offer important insight on the integrated strength of various fundamental mechanisms which govern the evolution of galaxies, such as stellar feedback, black hole activity, and mergers. 

One of the most robust ways of constraining the DM content within galaxies is through the measurement of galaxy kinematics. The rotation curves and pressure support of galaxies reflect their total density profile, providing an estimate of the DM content assuming that the baryonic content is known \citep{Rubin:1980aa, Burkert:1995aa, Burkert:1997aa,Persic:1996aa, McMillan:2017aa,Zhu:2023aa}. The fraction of DM to the total mass, $\fDM$, is a by-product of the mass assembly history of galaxies and of their current growth \citep[see][and references therein]{Wechsler:2018aa}.

The study of these density profiles within dwarf galaxies, defined as having masses of $\Mstar < 3\times10^9 \thinspace \Msun$, has proved particularly fruitful, revealing a discrepancy between $\Lambda$CDM-predicted and observed density profiles know as the core-cusp problem. Where collisionless (cold) DM-only cosmological simulations predict cuspy profiles, with densities increasing steeply at small radii, measurements of the rotation curves in dwarf galaxies in the local Universe show "core"-like profiles that flatten at central radii \citep{Moore:1994aa,Burkert:1995aa}. This core-cusp problem posed a direct challenge to $\Lambda$CDM , with propositions of different types of DM surfacing \citep{Hu:2000aa,Spergel:2000aa}. However, many studies have now shown that this problem can be alleviated by invoking strong baryonic feedback, likely from supernovae. This feedback, especially through repeated bursts of star formation, can flatten the DM cusp into a core \citep{Read:2005aa,Pontzen:2012aa,Pontzen:2014aa,Martizzi:2013aa}. This process has been observed in hydrodynamical simulations \citep[e.g.,][and references therein]{Chan:2015aa,Read:2016aa}, but details such as mass, redshift, and environmental dependence remain uncertain. Interestingly, cases of profiles cuspier than $\Lambda$CDM profiles have also been found \citep{Sonnenfeld:2012aa,Wang:2020ab,Li:2022ac}, which could be caused by adiabatic contraction of the DM halo due to high baryonic densities in the galaxy core \citep{Blumenthal:1986aa,Gnedin:2004aa,Li:2022ac}.

Another important relation that relates the potential of the DM halo to its host galaxy is the observed tight relation between stellar (or baryonic) mass and circular velocity, named the stellar (or baryonic) Tully-Fisher relation \citep[TFR;][]{Tully:1977aa}. The baryonic mass TFR (bTFR) is the more fundamental of the two, as it holds even down to low masses \citep{McGaugh:2000aa,McGaugh:2005aa}. Both TFRs have been found to hold for star-forming galaxies from $z\approx 0$ \citep{Reyes:2011aa,Lelli:2016aa} to cosmic noon \citep{Miller:2011aa,Ubler:2017aa}, albeit with an increase of scatter and a decrease of zero-point offset with redshift. Measuring the position of galaxies on the TFR plane at $z>4$ can provide constraints on the properties of the underlying haloes and their relative contribution to the total mass of the galaxy. 

The detailed mapping of rotation curves of galaxies in the local Universe has revealed that $\logMstar = 10-11$ star-forming galaxies are baryon-dominated in their central $\sim 1$ kpc, but DM dominated within their half-light, or effective, radii $\Reff$ \citep{Rubin:1985aa,Martinsson:2013aa}. On the other hand, early-type galaxies (ETGs) at $z\sim 0$ are heavily baryon-dominated within $\Reff$, suggesting a diverging mass assembly history \citep{Noordermeer:2007aa,Cappellari:2013aa,Serra:2016aa}. Interestingly, studies of DM fractions at cosmic noon ($z\approx 2$) in $\logMstar\gtrapprox 9.5$ galaxies found them to be similar to local ETGs, with baryon-dominated cores \citep{Price:2016uv, Wuyts:2016aa, Genzel:2017aa, Genzel:2020aa, Nestor-Shachar:2023aa}. \citet{Genzel:2020aa} interpret these findings as evidence for cored DM profiles, motivated by the rapid formation of massive haloes and galaxies at $z\sim 1-3$. In fact, cosmic noon marks the peak of the cosmic star-formation rate (SFR) density \citep{Madau:2014aa}, where gas accretion rates reached peak values \citep{Tacconi:2020aa} and stellar mass doubling scales were on the order of $t<0.4$ Gyr.

Measurement of the DM content of high-redshift galaxies ($z\geq4$) was made possible with the advent of the \textit{James Webb Space Telescope} (\textit{JWST}), whose NIRCam and NIRSpec instruments probe ionised gas emission lines out to $z\sim 10$. While kinematic studies had previously been done at such early times through cold gas measurements (i.e. [\ion{C}{II}] or CO) from ground based telescopes such as the Atacama Large Millimeter Array \citep[ALMA, e.g., ][]{Lelli:2021aa,Rizzo:2020aa,Rizzo:2021aa,Roman-Oliveira:2023aa,Pope:2023aa,Rowland:2024aa}, the lack of constraints on the stellar mass made estimating $\fDM$ difficult. Using kinematic modelling of \textit{JWST}/NIRSpec Multi Shutter Array (MSA) data \citep[see also][]{Saldana-Lopez:2025aa}, \citet{de-Graaff:2024ab, de-Graaff:2024aa} found that the five low-mass $\logMstar<9$ galaxies in their $z\sim6-8$ sample were DM dominated within $\Reff$ ($\fDM \sim 0.7$). By studying similar galaxies in the \textsc{tng50} simulations \citep{Pillepich:2019aa}, \citet{de-Graaff:2024aa} suggest that DM-dominated $\logMstar= 8-9$ galaxies at $z\sim 6$ are the progenitors of massive baryon-dominated systems at $z\sim2$. This finding predicts a DM dominated phase of early galaxy formation, with expectations that all low-mass galaxies have low baryon fractions in their central region.

When studying the DM fractions in the central regions of galaxies, it is also important to consider the effects of black holes (BHs) which could be hosted in their cores. Although BHs have been shown to have masses which are only a small fraction of the total dynamical mass \citep{Kormendy:2013aa}, recent \textit{JWST} studies have uncovered "overmassive" BHs in the early Universe, with $\Mbh/\Mstar \sim 0.1-0.01$ \citep{Harikane:2023aa,Kokorev:2023aa,Ubler:2023tn,Maiolino:2024aa,Juodzbalis:2025ab,Jones:2025aa}, even reaching $\Mbh/\Mstar > 2$ \citep{Juodzbalis:2025aa}. If the relatively low-mass galaxies at $z>4$ host massive black holes which experience accretion phases, then they could play a significant role in the galaxy's mass assembly history. Importantly, active galactic nuclei (AGN)-driven outflows can heavily disrupt the surrounding gas, particularly in low-mass galaxies \citep{Sijacki:2007aa,Nelson:2019aa,Koudmani:2021aa,Koudmani:2022aa,Carniani:2024aa}. The presence of such AGNs is important to study as it currently introduces more uncertainties in the interpretation of our kinematic measurements. 

With the synergy of NIRCam imaging and grism data, the kinematics of $\sim 200$ galaxies at $z=4-6$ were measured in \citet{Danhaive:2025aa} using a new forward-modelling Bayesian code, the Grism Emission-line Kinematics tOol \citep[\geko,][]{Danhaive:2025ac}\footnote{Available at \url{https://github.com/angelicalola-danhaive/geko}}. This sample is comprised of $\logMstar \approx7-10.5$ star-forming galaxies, with a wide range of rotational support $\rotsupp$. As shown in \citet{Danhaive:2025aa}, many of these galaxies have large dynamical masses compared to their stellar masses, pointing to large contributions from gas and DM. In this paper, we study the mass content of these galaxies in detail, placing the first statistical constraints on DM fractions at high redshift $z>4$. We briefly introduce our sample and methodology in Sec. \ref{sec:methods}, then present our dynamical mass measurements and discuss in the context of the Tully-Fisher (TFR) relation in Sec. \ref{sec:TFR}. We present our gas and DM fractions in Sec. \ref{sec:res-fractions}, and compare them to samples at lower redshift. In Sec. \ref{sec:discussion}, we interpret our measurements in the context of the shapes of DM density profiles, and explore potential contributions from BHs. Finally, we summarise our results in Sec. \ref{sec:conclusions}. Throughout this work, we assume $\Omega_0 =0.315$ and $H_0 = 67.4 \thinspace \text{km}\thinspace\text{s}^{-1}\thinspace\text{Mpc}^{-1}$ \citep{Planck-Collaboration:2020aa}.

\section{Observations, sample and methodology} \label{sec:methods}
In this section, we present the NIRCam grism and imaging data used in this work, along with a description of our sample (Sec. \ref{sec:sample}). We also describe the morpho-kinematic modelling of our sample (Sec. \ref{sec:modelling}), which is used to measure the rotational velocities and velocity dispersions of our galaxies. This sample, and the associated kinematic measurements, are described in \citet{Danhaive:2025aa}, so we only summarize them here but refer the reader to that work for more details. Finally, in Sec. \ref{sec:fraction-methods}, we detail our derivation of the gas and DM fractions for our sample.

\subsection{NIRCam data and sample selection} \label{sec:sample}

The analysis in this work uses both NIRCam imaging and grism spectroscopy. The NIRCam grism data is obtained from the FRESCO survey \citep[PI: Oesch, PID: 1895;][]{Oesch:2023aa,Covelo-Paz:2025aa} in GOODS-S and GOODS-N, and the CONGRESS survey (PIs: Egami \& Sun, PID: 3577; Sun et al. in prep) in GOODS-N. These surveys are comprised of grism data (R mode) in the F444W and F356W filters, respectively, probing \Ha\ emission at $z=3.8-6.5$. The data is reduced following \citet{Sun:2023ab} and \citet{Helton:2024aa} to obtain 2D spectra for each galaxy in the field of view. The 2D spectra are then continuum subtracted following the 2-step iterative technique described in \citet{Kashino:2023wv} to obtain 2D emission line maps for $\Ha$. 

In the regions of the FRESCO and CONGRESS surveys, there is a wealth of deep imaging data from the JADES survey \citep{Eisenstein:2023aa, Rieke:2023aa}. Specifically, the JADES survey has imaging in most of the NIRCam wide bands, as well as the F335M and F410M medium bands. In addition, the FRESCO survey obtained imaging in the F182M and F210M bands, and CONGRESS in F090W and F115W, complementing the existing JADES imaging data. For certain regions in our sample, we also have additional medium bands F182M, F210M, F430M, F460M, and F480M from the \textit{JWST }Extragalactic Medium Band Survey \citep[JEMS;][]{Williams:2023aa} in GOODS-S. The full details on the data reduction and generation of the drizzled mosaics and photometric catalogues can be found in \citet{Rieke:2023aa} and \citet{Robertson:2024aa}. This large array of photometric bands, in addition to the grism spectroscopic redshifts and emission line fluxes, allows us to fit the SEDs of all galaxies with \textsc{Prospector} \citep{Johnson:2021aa,Tacchella:2023aa,Simmonds:2024ab, Simmonds:2025aa}. We obtain measurements of the stellar masses ($\Mstar$), star-formation rates (SFRs), and star formation histories (SFHs) used in this work.

As mentioned above, the parent sample used in this work is described in \citet{Danhaive:2025aa}. This \Ha\ emitter sample is built on the photometric redshift estimates from \textsc{EAZY} \citep{Brammer:2008aa, Hainline:2024aa}, which are then confirmed by Gaussian fitting of the 1D grism spectrum followed by visual inspection \citep{Helton:2024aa, Lin:2025aa}. We select galaxies with an \Ha\ S/N cut of 10 (582 galaxies), from which we select our final sample of 213 galaxies based on additional S/N cuts and position angle (PA) cuts to ensure that galaxies are at an angle with respect to the grism dispersion direction. Specifically, the 213 galaxies are separated into three samples based on the robustness of the kinematic measurements, as described in \citet{Danhaive:2025aa}. In this paper, we only consider the 163 galaxies from the gold and silver samples, and we exclude the unresolved sample. The latter contains systems with sizes smaller than the full-width half-maximum (FWHM) of the F444W (or F356W) point-spread function (PSF), or with unresolved velocity gradients. However, for simplicity, we do not further distinguish between the gold and silver subsamples in this paper, treating the 163 galaxies together as our full final sample. 

\subsection{Morpho-kinematic modelling}\label{sec:modelling}

To obtain measurements of the rotational velocity and velocity dispersion, we use the Bayesian-inference fitting tool \geko\ \citep{Danhaive:2025ac} which forward models the grism data and compares it with the observed 2D spectrum to obtain the posterior distributions of the input morphological and kinematic model parameters. 

The \Ha\ emission line map is modelled with a \sersic profile \citep{Sersic:1968aa}:
\begin{equation}
    I(r) = I_{\rm e} \exp\Biggl( -b_n\left[ \left(\frac{r}{r_{\rm e}}\right)^{1/n}-1\right]\Biggr),
\label{eq:S\'{e}rsic_profile}
\end{equation}
where $I_{\rm e}$ is the intensity at the effective radius $r_{\rm e}$ and $n$ is the S\'{e}rsic index. We use the near-UV image from JADES, probed by the F150W filter at $\lambda_{\rm rest} \approx 2000 \angstrom$, as a prior for the morphology of the \Ha\ line. The rest-frame near-UV is chosen because it traces young stars responsible for ionizing the nearby gas and producing the \Ha\ emission. To model the F150W images, we use the Bayesian code \textsc{Pysersic} \citep{Pasha:2023aa}, which models the image with a PSF-convolved \sersic profile. The width of the priors is obtained by doubling the uncertainties obtained from the \textsc{Pysersic} modelling. 

The \Ha\ gas kinematics are modelled with an arctangent velocity curve \citep{Courteau:1997uw, Miller:2011aa}
\begin{equation}
    V_{\text{rot}}(r_{\text{int}}, r_{\text{t}}, V_{\rm a}) = \frac{2}{\pi}V_{\rm a}\arctan{\frac{r_{\text{int}}}{r_{\text{t}}}},
\end{equation}
where $V_{\text{rot}}$ is the rotational velocity at a given radius $r_{\text{int}}$ in the intrinsic galaxy plane, $V_{\rm a}$ is the asymptotic value to which the arctangent rotation curve converges to at large radii $r_{\text{int}} \rightarrow \infty$ and $r_{\text{t}}$ is the turn-around radius of the rotation curve. To project this velocity on the observation plane, we need to account for the galaxy's inclination $i$:
\begin{equation}
        V_{\text{obs}}(x,y) = V_{\text{rot}}(r_{\text{int}}, r_{\text{t}}, V_{\rm a})\cdot\sin{i}\cdot\cos{\phi_{\text{int}}},
\end{equation}
where $\phi_{\text{int}}$ is the polar angle coordinate in the galaxy plane. To compute the inclination, we assume an intrinsic axis ratio $q_0=0.2$ to account for the thickness of galaxies at high redshift \citep{Wuyts:2016aa, Genzel:2017aa, Price:2020wf,Ubler:2024aa}. We adopt a constant velocity dispersion across the galaxy. 

The model \Ha\ intrinsic map is convolved with the kinematics to form a model 3D cube, which is then convolved with the instrument PSF and line-spread function (LSF). Finally, using the grism dispersion function, the cube is projected onto the observed grism 2D space. The priors for the kinematic parameters are uniform. We place a constraint on the turn-around radius $r_{\text{t}}$ to be smaller than the effective radius $r_{\rm e}$ \citep{Miller:2011aa}. 

\subsection{Gas and DM fractions}\label{sec:fraction-methods}

In order to infer gas masses from our star-formation rates, we use the empirical relations calibrated in \cite{Tacconi:2020aa}. We assume here that the gas component is dominated by the molecular phase, whose mass is constrained in these relations. In dense, highly star-forming systems it is expected that the molecular phase dominates within $\Reff$, and the atomic phase only contributes to a minor degree \citep[e.g.][]{Leroy:2008aa,Saintonge:2017aa}. The ionised component is small in most systems \citep[e.g. see][and references therein]{Kennicutt:2012aa}. Given the small sizes ($\Reff \approx 0.5-1$ kpc, \citealt{Danhaive:2025ab}) and high specific star-formation rates ($\text{sSFR}$, with mass-doubling timescales of $t_{\rm double}\lessapprox 30$ Myr) of the galaxies in our sample, this is a valid assumption.

In these equations, the molecular gas mass is estimated using the sSFR of the system and the integrated depletion timescale for converting the gas into stars. This timescale depends on the redshift and therefore provides a more accurate conversion than fixed redshift relations \citep[e.g.][]{Kennicutt:1998vu}. Furthermore, the latter is calibrated to the local universe, where gas fractions are overall lower, whereas the \citet{Tacconi:2020aa} relations are calibrated from $z\sim 0 - 5.5$ from almost $2000$ objects or stacks \citep[see also][]{ Saintonge:2017aa,Tacconi:2018aa,Freundlich:2019aa}. Specifically, the gas-to-stellar mass ratio $\mu_{\text{gas}} = M_{\text{gas}}/M_\star$ scales with redshift $z$, sSFRs, and stellar mass \Mstar, where the scaling factors are computed using observational measurements of ionized and molecular gas from the literature. The relation for $\mu_{\text{gas}}$ is:

\begin{align*}
    \log \mu_{\text{gas}} =& ~A + B\times (\log(1+z) - F)^2 \\
    +& ~C\times \log (\text{sSFR}/\text{sSFR}_{\rm MS}(z,\Mstar)) \\
    +& ~D\times (\log \Mstar -10.7),
\end{align*}
with the star-forming main sequence (SFMS) defined as in \citet{Speagle:2014vq}, and the parameters $A,B,C,D$ and $F$ are the parameters obtained in \citealt{Tacconi:2020aa} (Table 2b) from error-weighed, multi-parameter regression. SFMSs such as \citet{Speagle:2014vq} can suffer from sample selection effects, especially on the low mass end, which can bias the inferred relation \citep[e.g.][]{McClymont:2025ab,Simmonds:2025aa}. We also note that the \citet{Kennicutt:1998vu} relation assumes solar metallicities, which can have significant impacts ($\sim 0.3$ dex) on the inferred SFRs, particularly for low-mass, metal-poor galaxies. 

Using the aforementioned scaling relation, we can then calculate gas fractions on the galaxy scale,
\begin{equation}
    f_{\text{gas}} = \frac{\mu_{\text{gas}}}{1+\mu_{\text{gas}}} = \frac{\Mgas}{\Mgas + \Mstar},
    \label{eq:fgas}
\end{equation}
and infer the total baryonic masses:
\begin{equation}
    M_{\text{bar}} = M_{\text{gas}} + \Mstar = (\mu_{\text{gas}} + 1)\Mstar.
\end{equation}

To quantify the uncertainties in our inferred gas fractions, we propagate the uncertainties from the inputs to the scaling relation ($\Mstar$ and SFR), along with the uncertainties of the constant parameters of the relation itself ($A,B,C,D$ and $F$). To assess the systematics introduced by our choice of gas mass estimate, we compare the $\fgas$ measurements obtained from the scaling relation to those inferred by a simple conversion
\begin{equation}
    \Mgas = \mathrm{SFR} \times t_{\rm dep}
\end{equation}
with a depletion time $t_{\rm dep} = 0.5-1$ Gyr. Although using this relation increases the intrinsic scatter, our median fractions remain consistent. If instead we change the SFMS used in the \citet{Tacconi:2020aa} relation, from \citet{Speagle:2014vq}, and adopt the one from \citet{Simmonds:2025aa}, we find that our inferred gas fractions are on average higher by $\Delta\fgas \approx 0.05-0.1$. We note that in both of these cases, the change in $\fgas$ does not visibly affect the median values of our inferred dark-matter fractions. However, it is important to note these uncertainties in the gas fractions, which can only be addressed with constraints from direct observations of the molecular gas content. 

In virial equilibrium, the circular velocity of galaxies at a radius $r$ can be related to the combined effects of gravity and turbulence-induced pressure. The former is reflected in the rotational velocity, and the latter is an asymmetric drift correction to account for the pressure support \citep{Burkert:2010aa,Newman:2013aa, Wuyts:2016aa}. For an exponential disk, the circular velocity can be written as: 
\begin{equation}
     v_{\text{circ}}(r) = \sqrt{v_{\text{rot}}^2(r) + 2(r/r_{\rm s})\sigma_0^2}.
    \label{eq:v_circ}
\end{equation}
The circular velocity reflects the gravitational potential of the galaxy and is hence directly related to the dynamical mass,
\begin{equation}
    M_{\text{dyn}} = k_{\text{tot}}\frac{r_{\rm e}v^2_{\text{circ}}(r_{\rm e})}{G},
    \label{eq:dyn-mass}
\end{equation}
where $G$ is the gravitational constant and $k_{\text{tot}}$ is the virial coefficient \citep{Price:2020wf}. The virial coefficient $k_{\text{tot}}$ allows us to infer the total dynamical masses based on measurements out to $r_{\rm e}$. If we instead compute $\vcirc$ using a generalised \sersic\ profile (rather than an exponential disk, $n=1$), our inferred dynamical masses remain consistent within $\Delta(\log M_{\rm dyn})\lesssim 0.07$ (with median $\Delta(\log M_{\rm dyn})\lesssim 0.02$). We present results for the exponential disk assumption to provide a more direct comparison to other works in the literature. Because we have modelled our galaxies with $q_0=0.2$, we choose $k_{\text{tot}}=1.8$ as it is the coefficient for galaxies with $q_0=0.2$ and $n\sim 1-4$ \citep{Price:2022aa}. We note that varying $q_0$ in the interval  $q_0 = 0.15-0.25$  shifts $\sin i$ by $\Delta(\sin i)\lesssim0.05$, propagating to $\Delta(\log M_{\rm dyn})\lesssim0.05$. For more extreme values $q_0 = 0.50$, this correction can reach $\Delta(\log M_{\rm dyn})\approx 0.1$. 

Finally, using our dynamical mass measurements, we can compute the DM fractions:
\begin{equation}
    f_{\text{DM}} = \frac{ M_{\text{DM}}}{ M_{\text{dyn}}} = \frac{ M_{\text{dyn}} -  M_{\text{bar}}}{ M_{\text{dyn}}}.
    \label{eq:fdm}
\end{equation}
All our inferred fractions are calculated within the effective radius $\Reff$ since that is where our kinematic measurements are best constrained. We find that $20\%$ of our sample has negative and, therefore, unphysical DM fractions. This fraction drops to $18\%$ ($8\%$) when a deviation from $\fDM = 0$ is required at the level of $\sigma$ (3$\sigma$). These numbers are comparable to those reported in other studies \citep[e.g.,][]{Wuyts:2016aa}, and suggest inconsistencies in the modelling of the SEDs and the kinematics, including for instance positive age gradients \citep{DEugenio:2021aa,van-Houdt:2021aa} or non-equilibrium configurations. Although we discuss these systems in the text, unless otherwise mentioned we only consider systems with $\fDM>0$ for our analyses. 

\section{Dynamical masses \& Tully-Fisher relation} \label{sec:TFR}
In this section, we present the dynamical masses of our systems based on our kinematic modelling (Sec. \ref{sec:dynmass}), and discuss them within the context of the Tully-Fisher relation at high redshift (Sec. \ref{sec:sub-tfr}).

\subsection{Dynamical masses} \label{sec:dynmass}
\begin{figure}
    \centering
    \includegraphics[width=1\linewidth]{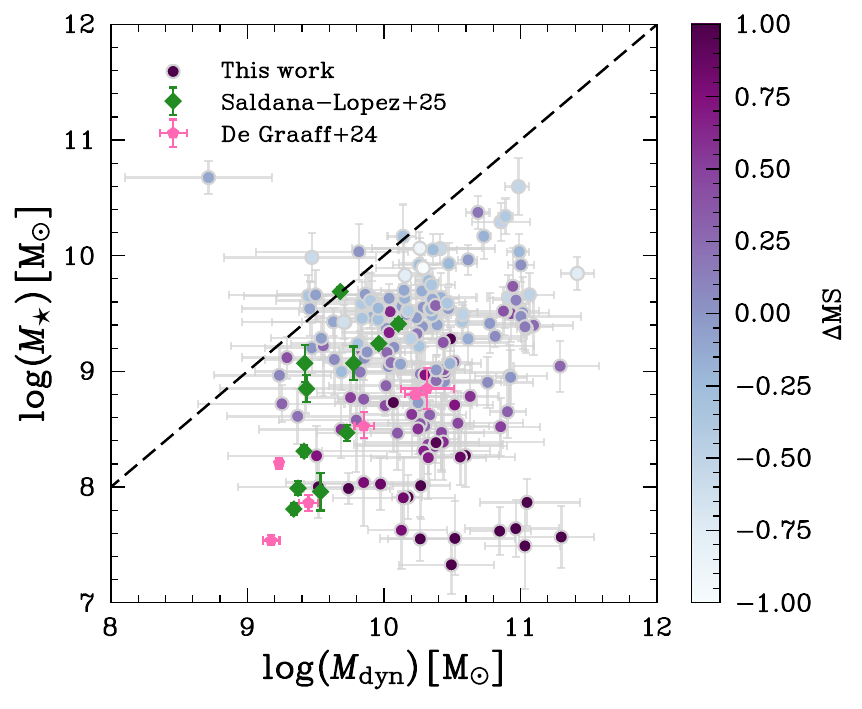}
    \caption{Comparison of the dynamical masses (\Mdyn) inferred from our modelling of grism data and the stellar masses ($M_{\star}$) inferred from SED fitting. We colour-code our galaxies based on their offset from the main sequence ($\Delta\rm MS$) as defined by \citet{Simmonds:2025aa}. The majority of our systems lie below the one-to-one relation (dashed line), consistent with a significant contribution to the dynamical mass from gas and/or DM. Six galaxies lie on the relation, highlighting discrepancies in the different mass estimates. We compare our values to ionised gas measurements from \citet{de-Graaff:2024ab} and \citet{Saldana-Lopez:2025aa} at similar redshifts.}
    \label{fig:mdyn-mstar}
\end{figure}

We present our results for the dynamical masses of our galaxies in Fig. \ref{fig:mdyn-mstar}. We find high dynamical masses spanning $\logMdyn = 9-11$, with the overall expected trend of increasing $\Mdyn$ with $\Mstar$, reflecting the baryons tracing the underlying gravitational potential. We find a significant scatter in the dynamical masses at fixed stellar mass, indicative of varying gas and DM contents. To investigate the source of this scatter, we colour-code our galaxies by their offset from the MS ($\deltaMS$), where we adopt the MS from \citet{Simmonds:2025aa}. We find that galaxies above the MS, which are also typically at lower masses, seem most distant from the 1-1 line, implying a lower stellar contribution to the baryonic and total mass.

We expect the dynamical masses to lie above the stellar masses, as they should also incorporate the gas and DM. We find consistent measurements $M_{\star}<M_{\text{dyn}}$ for the majority of the galaxies in our sample. We discuss the few systems showing unphysical $\Mdyn-\Mstar$ values in more detail in \citet{Danhaive:2025aa}, the discrepancy pointing to problems in the modelling of either or both masses. We compare our sample with measurements from \citet{Saldana-Lopez:2025aa} and \citet{de-Graaff:2024aa} at similar redshifts, and find that they probe the same regions of the parameter space. When combined, these two samples highlight a similar scatter of $\Mstar$ at fixed $\Mdyn$ as found in our sample. Overall, these dynamical masses are the key observable that allows us to derive DM fractions (Sec. \ref{sec:methods}). We note that in a recent study, \citet{Phillips:2025aa} compare intrinsic kinematic measurements from simulations to mock NIRSpec/IFU fits, and find that despite biases in the recovered values of $\vrot$ and $\disp$, the dynamical masses remain relatively robust. 

\subsection{The Tully-Fisher relation (TFR)}\label{sec:sub-tfr}

Local star-forming disk galaxies exhibit a tight relationship between their stellar mass and circular velocity ($\vcirc$, Eq. \ref{eq:v_circ}), namely the TFR relation. This relation is typically parametrised as a power law, $M\propto V^{a}$, or 

\begin{equation}
    \log \Mstar = a \cdot \log \vcirc + b,
    \label{eq:tfr}
\end{equation}
where $a$ is the slope and $b$ is the zero-point offset, and $\Mstar$ is in units of $\Msun$ and $\vcirc$ in km/s. This relation is analogous to the $\Mstar-\Mdyn$ relation from Fig. \ref{fig:mdyn-mstar}, given the fundamental link between circular velocities and dynamical masses (Eq. \ref{eq:dyn-mass}) when assuming virial equilibrium. By understanding where our galaxies lie on the TFR plane, we can place constraints on their mass content compared to galaxies at lower redshift.

In order to explore the offset between our sample at $z=4-6$ and the sTFR at cosmic noon and in the local Universe, we fit Eq. \ref{eq:tfr} to our sample with the fixed slope from \citet{Reyes:2011aa} at $z\sim 0$ ($a = 3.60$). For our fit, we use the Bayesian inference package \textsc{emcee} \citep{Foreman-Mackey:2013aa}. \textsc{emcee} samples the parameter space using an implementation of the affine invariant ensemble sampler for Markov chain Monte Carlo \citep{Goodman:2010aa}, resulting in posterior distributions for the model parameters. Importantly, it allows us to self-consistently fit for the intrinsic scatter $\sigma_{\rm int}$ around the relation, assuming that our adopted uncertainties for the various measurements are reliable. We discuss this assumption in Sec. \ref{sec:caveat-mass}. For the \textsc{emcee} fitting, we assume uniform priors for all of the parameters and conduct the MCMC sampling with 50 walkers, a 2000 step burn-in phase, and 10,000 samples. We thin our chains to remove autocorrelation between samples, checking for convergence in the trace plots and in the effective number of independent samples. 

\begin{figure}
    \centering
    \includegraphics[width=1\linewidth]{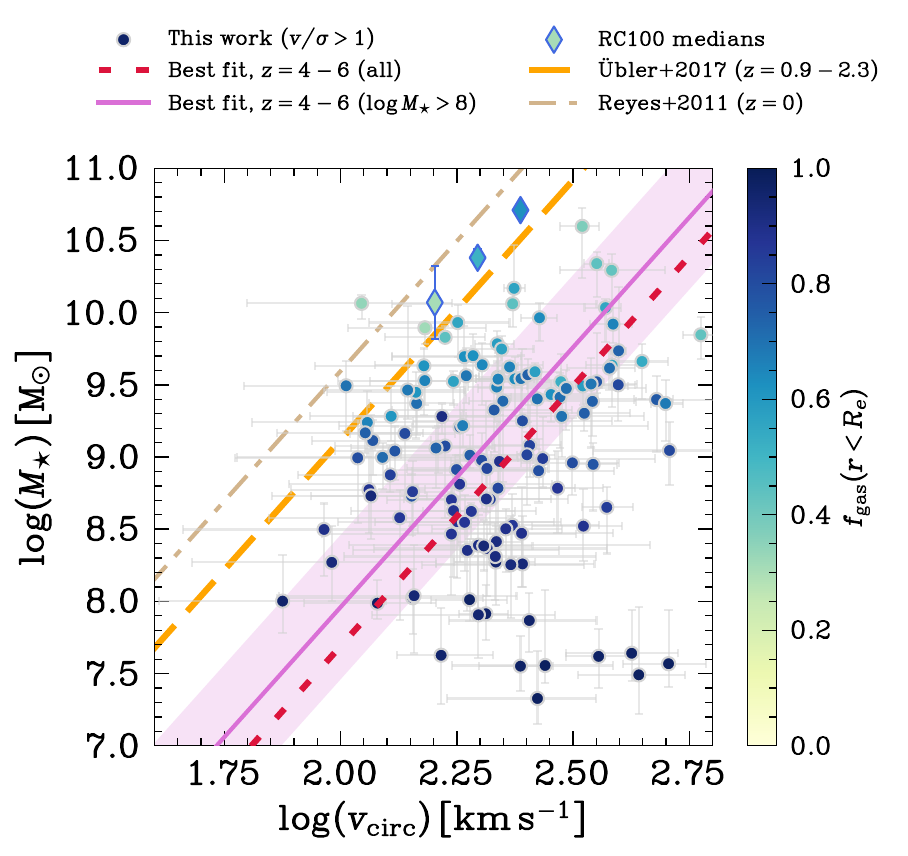}
    \caption{The stellar Tully-Fisher plane for our sample, where the best-fit linear relation for galaxies with $\logMstar>8$ is shown in solid purple (dotted red for the full sample). The circular velocity ($\vcirc$) is evaluated at $\Reff$ with an asymmetric‑drift correction. We find that most of our galaxies lie along the relation, albeit with a large scatter, but the lowest-mass galaxies drop off. We compare our measurements at $z\sim 5$ to the relation found for the KMOS$^{\rm 3D}$ $z=0.9-2.3$ star-forming galaxies from \protect\cite{Ubler:2017aa} (dashed orange line), and the medians from the RC100 sample \citep[diamonds, ][]{Nestor-Shachar:2023aa}. We also include the TFR at $z\sim 0$ from \citet{Reyes:2011aa} for reference (dot-dashed tan curve). We fit our data with the same fixed slope as these works, $a = 3.60$, but computing our own zero-point $b$. We find $\Delta b \approx -1.3$ dex between from  $z=0.9-2.3$ to $z=4-6$. The large intrinsic scatter around the relation ($\sigma_{\rm int} = 0.49 \pm 0.06$; shaded region) indicates that it only beginning to emerge at $z\sim5$.}
    \label{fig:TFR}
\end{figure}

We find the following best fit parameters for the $\logMstar>8$ sample, where we are more complete:
\begin{align}
    a = & ~3.60 \rm \thinspace \thinspace (fixed) \\
    b = & ~0.76 \pm 0.07 \\
    \sigma_{\rm int} =& ~ 0.49 \pm 0.06 ~.
\end{align}
We plot the results for this fiducial fit at $z\sim 5$ in purple in Fig. \ref{fig:TFR}, and also include a fit for the whole sample in red. The large intrinsic scatter, shown by the shaded regions, reflects that the $z\sim5$ galaxies in our sample are not settled around the TFR. They are likely undergoing kinematic and/or morphological changes on relatively short timescales, which influence the measured circular velocities at fixed stellar masses. More generally, this implies that galaxies have not yet settled into more stable rotating disks, where the TFR is expected to hold \citep[e.g.,][]{Ubler:2017aa}.

Despite the lack of a tight TFR in our sample, we explore the evolution of the zero-point offset $b$ from $z\sim 0$ \citep{Reyes:2011aa} and $z\sim 1-3$ (\citealt{Ubler:2017aa}, see also \citet{Tiley:2016aa}) to $z\approx4-6$. Both works adopt the same slope. We find a significant increase in the zero-point offset, $\Delta b \approx -1.3$ dex, between our TFR and the one measured at cosmic noon \citep{Ubler:2017aa}. This evolution is more significant than the $\Delta b \approx -0.4$ dex evolution from $z\sim 0$ to cosmic noon. The observed decrease of $b$ with redshift could have (a mixture of) different origins. To visually compare with galaxies at cosmic noon, we plot on Fig. \ref{fig:TFR} the RC100 sample from \citet{Nestor-Shachar:2023aa} at $z=0.6-2.5$. It is clear that these cosmic noon galaxies occupy a different parameter space in the TFR plane, having higher stellar masses ($\logMstar \approx 10-11$) and lower gas fractions ($\fgas\lessapprox 0.5$) than our sample. 

First, an increase of gas fractions naturally explains the shift of the TFR to higher circular velocities (at fixed mass). This is because $\vcirc$ reflects the total mass of the galaxy, both baryons and DM, so if the fraction of stars with respect to the other components decreases, the TFR will shift. In the same way, an increase in the DM fractions can also contribute to this shift. It is also possible that $b$ has a mass dependence, since our sample probes smaller masses than the KMOS$^{\rm 3D}$ and RC100 samples. It is expected that gas fractions increase at low stellar masses, naturally biasing the stellar TFR. A larger sample spanning larger redshift and mass ranges would be needed to assess the mass dependence of $b$. 

In light of this, we also explore the bTFR in our sample, but similarly to the sTFR we do not find a tight relation akin to what has been measured at cosmic noon and the local Universe \citep{McGaugh:2012aa,Lelli:2016aa}. We also find a similar offset towards higher circular velocities at fixed baryonic mass. Together, these points suggests that DM contributes significantly to the total mass of the galaxies in our sample. This is in contrast to the baryon-dominated galaxies seen at cosmic noon. To correctly measure the bTFR, gas masses would need to be directly measured, which is beyond the scope of this work and our available data.

In terms of intrinsic scatter ($\sigma_{\rm int} = 0.49 \pm 0.06$), we find that it is roughly twice that reported at cosmic noon \citep[$\sigma_{\rm int} = 0.22$,][]{Ubler:2017aa}, where they also find an increase with respect to the local Universe \citep[$\sigma_{\rm int} = 0.10-0.13$][]{Reyes:2011aa}. As discussed in \citet{Ubler:2017aa}, one reason for this could be the fact that galaxies at high-redshift are less settled \citep{Simons:2016aa}, and are seen more often in non-equilibrium states \citep{Covington:2010aa}. Also, we rely on 2D grism data in this work and, hence, suffer from more uncertainties than studies with 3D kinematic data. Furthermore, galaxies at high redshift are smaller and lower mass on average, which makes their kinematics more difficult to constrain. 

Finally, we note that the lowest-mass galaxies in our sample, with $\logMstar < 8$, distinctively fall off our best-fit relation, with high circular velocities of $\vcirc>150$ km/s despite their low masses. We also fit our full sample, including these $\logMstar<8$ objects, and find a smaller zero-point ($b = 0.59 \pm 0.08$) and a larger intrinsic scatter ($ \sigma_{\rm int} = 0.73 \pm 0.07$). These galaxies also have gas fractions close to 1, and are most likely undergoing starbursts ($\Delta\rm MS >0$). It is possible that the circular velocities measured for these systems have a non-negligible contribution from non-circular motions, such as radial inflows and/or outflows of gas. This would also contribute to the large observed scatter. Due to their low stellar masses and, hence, small spatial extent, such effects would be difficult to detect in our modelling. Our sample is in general not complete, so our discussion regarding the TFR aims at comparing our sample's physical properties with those of galaxies at lower redshift and not fully defining the TFR at $z=4-6$.

\section{The gas and DM content at $z>5$}\label{sec:res-fractions}

In this section, we quantify the baryonic and DM content of galaxies at $z=4-6$, using gas and DM fractions derived from our kinematic modelling (Sec. \ref{sec:methods}). We first study the dependence of these fractions on stellar mass (Sec. \ref{sec:fracs-mass}), then we focus on the relation between DM fractions and baryonic surface density in Sec. \ref{sec:sigmabar}. Finally, we put our findings in the context of works at lower redshift to discuss the redshift evolution of DM fractions (Sec. \ref{sec:z-evol}).

\subsection{Baryon content at $z\sim 5$}\label{sec:fracs-mass}

\begin{figure}
    \centering
    \includegraphics[width=1\linewidth]{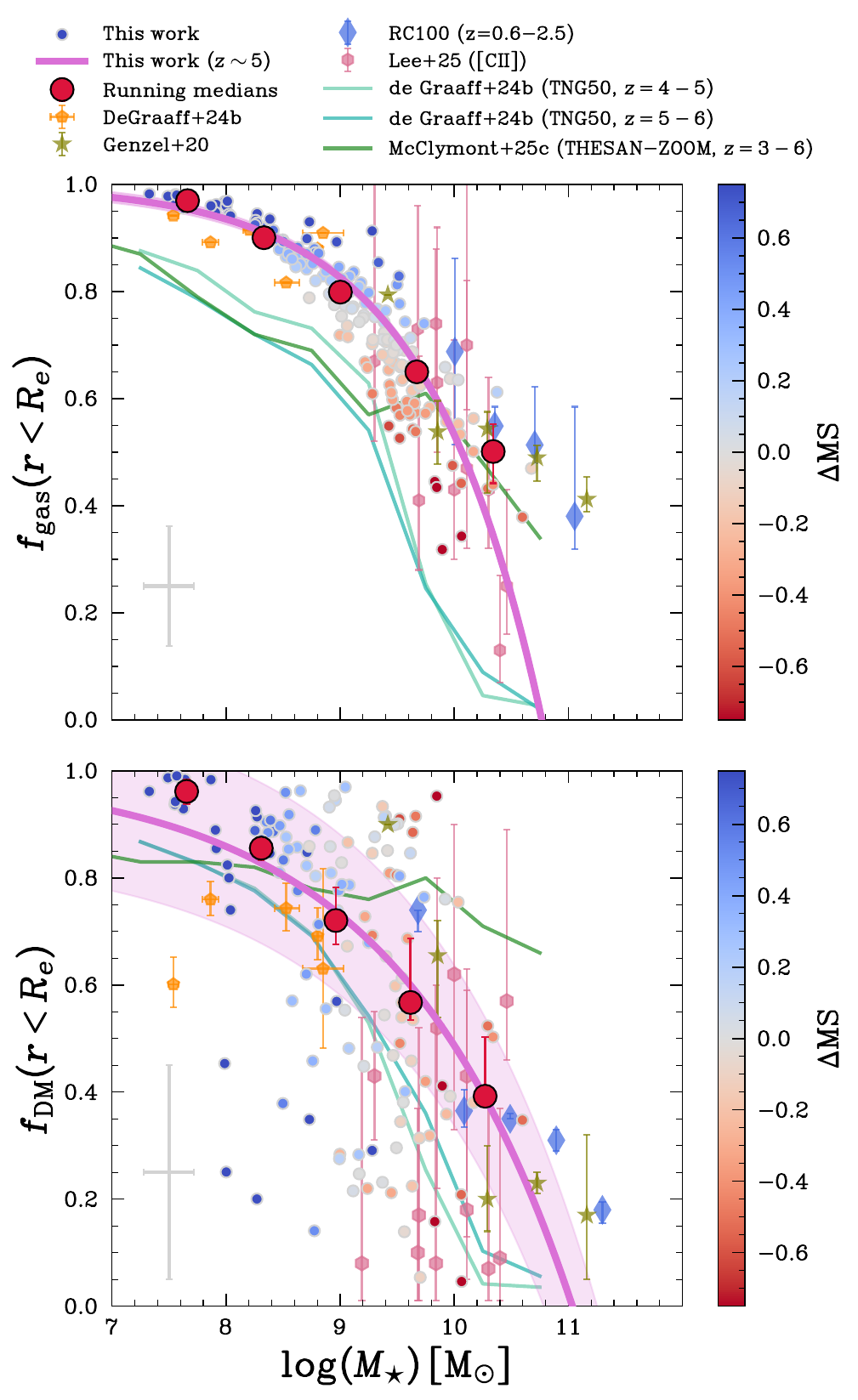}
    \caption{Gas (top) and DM (bottom) fractions (within $\Reff$) as a function of stellar mass, with the characteristic uncertainty shown in grey. The $f_{\text{gas}} (r<\Reff)$ (Eq. \ref{eq:fgas}) is computed using the sSFR, redshift, and stellar mass of each galaxy following \protect\cite{Tacconi:2020aa}. The $f_{\text{DM}} (r<\Reff)$ is computed using $M_{\text{gas}}$ along with $M_{\text{dyn}}$ and $M_{\star}$ as shown in Eq. \ref{eq:fdm}. We fit our data with a power-law (solid purple line) for both relations, showing the intrinsic scatter (shaded region) only for $\fDM$. We cannot constrain $\sigma_{\rm int}$ for $\fgas$ due to the inferred uncertainties being significantly larger than the scatter in the outputs of the \citet{Tacconi:2020aa} scaling relation for our sample (see Sec. \ref{sec:gas-fracs}). We compare our results with measurements from \citet{Genzel:2020aa} (green stars) and \citet{Nestor-Shachar:2023aa} (RC100; blue diamonds) at cosmic noon and \citet{de-Graaff:2024ab,de-Graaff:2024aa} (orange pentagons) and \citet{Lee:2025aa} (pink hexagons) at high redshift. For most of our sample, we find high gas and DM fractions $f > 0.5$. We find relatively good agreement with median trends from the \textsc{thesan-zoom} simulations \citep[green solid line;][]{McClymont:2025ac} and the \textsc{tng50} simulations \citep[blue solid lines;][]{de-Graaff:2024aa}.}
    \label{fig:fractions-mass}
\end{figure}

Fig. \ref{fig:fractions-mass} shows the inferred gas ($f_{\text{gas}}$) and DM ($f_{\text{DM}}$) fractions as a function of stellar mass for the galaxies in our samples. We note that $f_{\text{gas}}$ is defined relative to the baryonic mass (stars + gas), whereas $f_{\text{DM}}$ is defined relative to the total mass (DM + baryons). We fit both relations with a power law of the form
\begin{equation}
    f(z,M_{\star}) = 1 - 10^{\alpha (\log M_{\star}-9) + \beta + \gamma(\frac{1+z}{6})},
    \label{eq:power-law}
\end{equation}
where $\alpha$ and $\gamma$ parametrize the mass and redshift dependence, respectively, and $\beta$ is the normalisation at $\logMstar = 9$ and $z=5$. The power-law shape is consistent with the equation from \citet{Tacconi:2020aa} for the gas fractions. Similarly to the TFR, these fits are done with \textsc{emcee} using the same general setup (see Sec. \ref{sec:TFR}). We summarize the best-fit parameters for both the gas and the DM fractions in Tab. \ref{tab:best-fit-fracs}. For the gas fractions, we are not able to constrain the intrinsic scatter ($\sigma_{\rm int}$) around the best-fit relation. As explained in Sec. \ref{sec:methods}, we compute the uncertainties on $\fgas$ by propagating the uncertainties on both our input parameters and the constant parameters of the \citet{Tacconi:2020aa} scaling relation. This yields relatively large uncertainties of the order of $\Delta \fgas \approx 0.1$ (see characteristic error in Fig. \ref{fig:fractions-mass}). On the other hand, the outputs $\fgas$ from the scaling relation vary only a small amount at fixed mass. This is in part due to the small range probed in terms of $\deltaMS$ in our sample, since we are not sensitive to galaxies well below the MS. Because of this, our uncertainties are significantly larger than the scatter shown by our inferred $\fgas$, meaning that we cannot constrain $\sigma_{\rm int}$.

\renewcommand{\arraystretch}{1.5} % Increase row spacing
\setlength{\tabcolsep}{3pt} % Slightly reduce column spacing

\begin{table}
\centering
\begin{tabular}{c|c|c}
Parameter & $f_{\text{gas}}$ & $f_{\text{DM}}$ \\
\hline
$\alpha$ & $0.43\pm 0.03$ & $0.28\pm 0.05$ \\
\hline 
$\beta$ & $-0.73\pm 0.03$ & $-0.57\pm 0.04$ \\
\hline 
$\gamma$ & $-0.36\pm 0.45$ & $0.34\pm 0.65$ \\
\hline
$\sigma_{\rm int}$ & -- &$0.15 \pm 0.02$\\
\end{tabular}
\caption{Summary of the parameters for the power-law fit (Eq. \ref{eq:power-law}) to the $f_{\text{gas}}-$\Mstar\ and $f_{\text{DM}}-$\Mstar\ relations (Fig. \ref{fig:fractions-mass}). We cannot constrain $\sigma_{\rm int}$ for the $\fgas-\Mstar$ relation due to the inferred uncertainties on $\fgas$ being significantly larger than the scatter in the outputs of the \citet{Tacconi:2020aa} scaling relation for our sample (see Sec. \ref{sec:gas-fracs} for more details).}
\label{tab:best-fit-fracs}
\end{table}

\subsubsection{Gas fractions} \label{sec:gas-fracs}

As described in Sec. \ref{sec:methods}, we cannot measure gas masses directly, so we infer them based on the galaxies' SFRs, stellar masses, and redshifts through the \citet{Tacconi:2020aa} scaling relation. As seen in the top panel of Fig. \ref{fig:fractions-mass}, our galaxies have high predicted gas fractions $\fgas = \Mgas/\Mbar$, with almost all galaxies having $\fgas>0.5$. These systems have more gas than stars in their central region $r<\Reff$. These high gas fractions are driven in part by the high SFRs, since gas must accrete onto the galaxy to cool and fuel star formation. Gas fractions are also expected to be overall higher at high redshift, but we cannot constrain the redshift dependence of $\fgas$  due to the small redshift range probed in this work. However, as expected, our inferred dependence of $\gamma = -0.36 \pm 0.45$ is consistent with its counter-part in the \citet{Tacconi:2020aa} scaling relation ($D = -0.41 \pm 0.03$), albeit with very large uncertainties.

We compare our gas fractions with estimates at cosmic noon from \citet{Genzel:2020aa} and \citet{Nestor-Shachar:2023aa}, which show a large scatter at $\logMstar \approx 10-11$. Because there is little overlap between this mass range and the one probed in this work, a direct comparison to quantify the independent effects of redshift is not possible. However, their results are consistent with a decrease, on average, of gas fractions with stellar mass.

The high gas fractions at high-redshift are consistent with other predictions of the increase of gas fractions with redshift \citep{Cresci:2009aa, Genzel:2015aa, Pillepich:2019aa}. However, this effect appears to also have a strong mass component. The gas fractions found in \citet{de-Graaff:2024ab, de-Graaff:2024aa} at $z\approx 6- 8$ are consistent with our results. In fact, although \citet{de-Graaff:2024ab} use the local relation from \citet{Kennicutt:1998vu} to infer gas masses from SFRs, their measurements lie nicely along our best-fit relation. This suggests that this local relation holds well at $z\sim6$, at least at the low masses ($\logMstar<9$) probed by their work. The lower gas fractions, derived from [CII] detections, presented in \citet{Lee:2025aa} at $z\sim 4-6$ are consistent with our best-fit power-law. 

The steep mass dependence of $\fgas$ on $\Mstar$ is also seen in the TNG50 simulations \citep{Pillepich:2019aa, de-Graaff:2024aa} and the \textsc{thesan-zoom} simulations \citep{Kannan:2025aa,McClymont:2025ac}. At low masses, we find higher gas fractions on average than both of these simulations. This is partly because the relations from \citet{Tacconi:2020aa} and \citet{Speagle:2014vq} do not reach the low mass ($\logMstar<9$) end, and hence the extrapolation could cause an over-estimate of gas fractions at low masses. Importantly, our sample is biased to very star-forming galaxies at the low-mass end, which naturally explains our higher medians.

\subsubsection{DM fractions}

We now discuss the bottom panel of Fig. \ref{fig:fractions-mass}, where we plot our DM fractions (Eq. \ref{eq:fdm}) as a function of stellar mass. These fractions are computed within the half-light, or effective, radius $\Reff$ of the \Ha\ emission to avoid extrapolations of our kinematic measurements to larger radii. This implies that the fractions discussed in this work are relevant for the central $r=1.20\pm 0.05$ kpc of each galaxy \citep[sample median,][]{Danhaive:2025ab}. We see a large scatter in the DM content of galaxies at fixed stellar mass, but nonetheless find that low-mass galaxies ($\logMstar <9$) are predominantly DM dominated with $f_{\text{DM}}(r<\Reff) > 0.5$. Above $\logMstar =9$, we observe a larger scatter and an overall decline in DM fractions with stellar mass, as is highlighted by the running medians and our best-fit power-law in Fig. \ref{fig:fractions-mass}.  We colour-code our galaxies by their offset from the main sequence, but we do not find a clear trend.

We put our $\fDM$ measurements in the context of works at cosmic noon, which probe a higher mass range $\logMstar \approx 10 -11$. As shown in Fig. \ref{sec:fracs-mass}, \citet{Genzel:2020aa} find that star-forming galaxies at cosmic noon have baryon-dominated cores, with $\fDM\lessapprox0.5$. This is consistent with the extrapolation of our power-law fit to these high masses, suggesting that the mass dependence of $\fDM$ exists independently of redshift. In fact, \citet{de-Graaff:2024aa} show with the TNG simulations that the baryon fraction within 1 kpc, $f_{\text{bar}}(r<1$ kpc), remains constant with redshift at fixed mass. However, it is important to note that our measurements are not at fixed apertures, and we need to take into account the redshift evolution of galaxy sizes \citep{Shibuya:2015aa,Ward:2024aa,Danhaive:2025ab,Allen:2025aa,Yang:2025aa}. Nonetheless, our findings suggest that the high DM fractions seen at high-$z$, both in this work and \citet{de-Graaff:2024aa}, are at least in part driven by the lower stellar masses probed compared to lower redshifts. In fact, \citet{Lee:2025aa} probe higher stellar masses on average ($\logMstar>9$) and find DM fractions ranging from $\fDM\approx 0.1$ to $\fDM\approx 0.6$, consistent with the scatter seen in our larger sample at those masses.

We also compare our DM fractions with predictions from the \textsc{thesan-zoom} simulations at $z=3-6$. In general, the simulations are consistent with the majority ($60\%$) of our sample having high DM fractions $\fDM>0.5$. However, \citet{McClymont:2025ac} find a weaker mass trend, which does not reproduce the population of $\logMstar>9$ galaxies with lower $\fDM$. This could point to mechanisms which drive DM out from the central regions, such as intense feedback from star formation or black holes (see Sec. \ref{sec:discussion} for more discussion). Within the simulations, this could also be due to a decrease in the number of galaxies at the high-mass end. Our observed dependence of $\fDM$ on $\Mstar$ is instead more consistent with the trend observed in the \textsc{tng50} simulations \citep{de-Graaff:2024aa}.

Some objects have high DM fractions close to $\fDM = 1$. Although these could be caused by an over-estimate of the dynamical mass, they can also point to steepening of the DM halo density profile in the central regions of these galaxies. We explore this in Sec. \ref{sec:alpha}. We also find that the inferred $f_{\text{DM}}$ fractions for 20\% of our sample are unphysical, i.e., $f_{\text{DM}} <0$. Negative fractions imply that the baryonic mass is larger than the dynamical mass, which could be caused by one or more of the following. The stellar mass inferred from SED fitting could be over-estimated (e.g. due to outshining effects, Sec. \ref{sec:caveat-mass}), the dynamical mass from the kinematics fitting could be under-estimated, or the gas fractions inferred from the \citet{Tacconi:2020aa} empirical relation could be over-estimated. Many galaxies in our sample, despite having good fits with low residuals, have clumpy morphologies, which could indicate clumpy SF and/or final stages of a merger. Irregular morphologies can bias the measurement of the rotational velocity due to the simplistic nature of the rotation curve model. Also, if a merger is in fact taking place, the stellar mass may be over-estimated due to the multiple-system nature of the object. In some of the systems, the \textsc{Prospector} posteriors for the stellar mass are distinctively double peaked, with a lower mass solution that would boost the $f_{\text{DM}}$. Finally, based on the \textsc{Prospector}-inferred SFHs, many of the systems are either in a burst or just went through a burst, with a peak in SF over a $\sim 5-10$ Myr timescale. Especially at low masses, these starbursts can disrupt the ordered rotation of the gas, leading to low and/or bias measurements of $v_{\text{rot}}$. 

\subsection{Baryonic surface density}\label{sec:sigmabar}

\begin{figure*}
    \centering
    \includegraphics[width=1\linewidth]{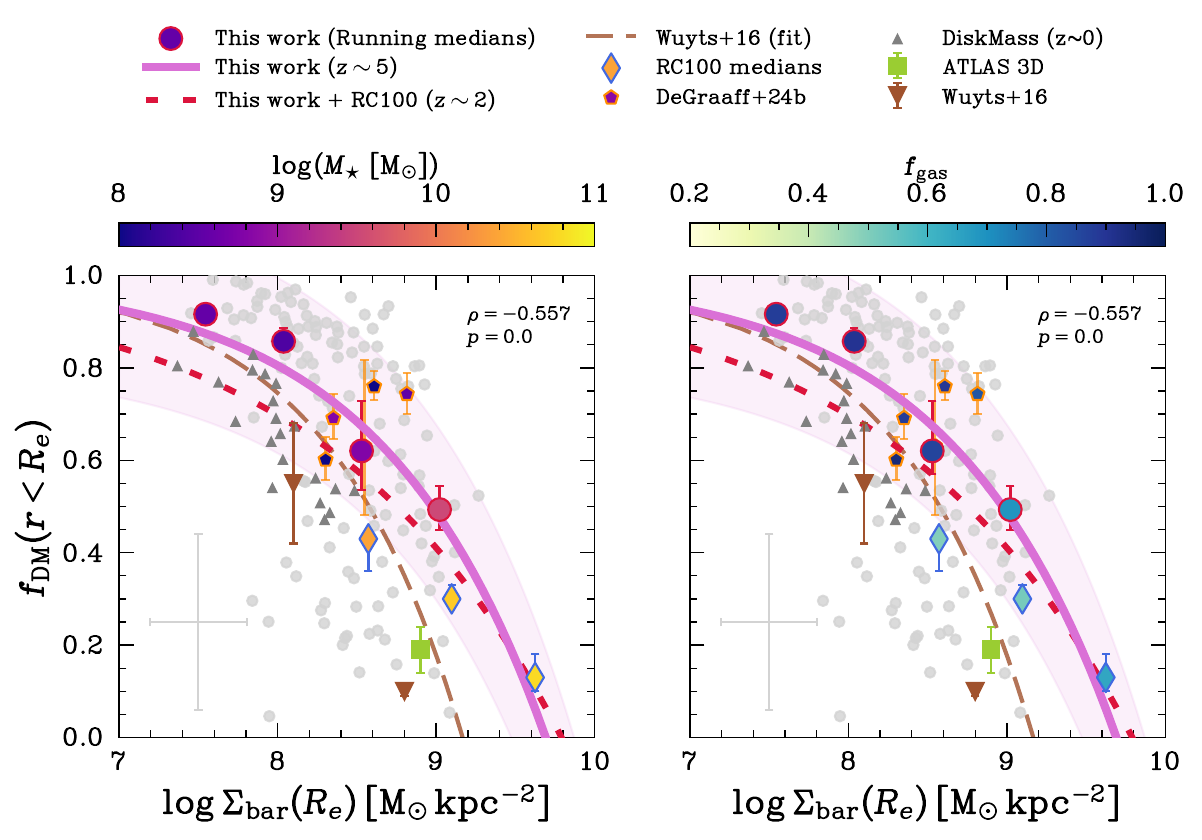}
    \caption{DM fraction as a function of baryonic surface density ($\Sigma_{\rm bar}$) for our sample (circles) and samples at lower \citep[diamonds, ][]{Nestor-Shachar:2023aa} and similar to higher \citep[pentagons, ][]{de-Graaff:2024aa} redshift. We colour-code points by stellar mass (left panel) and gas fractions (right panel). We show the characteristic uncertainty for our sample in grey. We find a strong negative correlation between $\fDM$ and $\Sigma_{\rm bar}$ ($\rho = -0.551, p < 0.001$), with a best-fit relation (solid purple line) steeper than \citet{Wuyts:2016aa} at cosmic noon (dashed brown line), and with an offset. We also plot the best-fit relation obtain when fitting our points with those from \citet{Nestor-Shachar:2023aa} (dashed-dotted red line). For comparison, we plot medians from ETGs at $z\sim 0$ from the ATLAS3D survey \citep[purple square; ][]{Cappellari:2013aa} and LTGs from the DiskMass Survey \citep[grey triangles; ][]{Martinsson:2013aa}. Our galaxies agree with \citet{Nestor-Shachar:2023aa} at low fractions $\fDM<0.5$, but lie above the relation at high fractions. The galaxies in our sample are smaller, less massive, and more gas rich than those probed at cosmic noon, driving the apparent shift in the relation. }
    \label{fig:fdm-sigmabar}
\end{figure*}

In order to investigate the origin of the DM fractions measured in our $z=4-6$ sample, we study their correlation with the baryonic surface density $\sigmabar$, computed within the effective radius of the \Ha\ emission, $\Reff$:

\begin{equation}
    \sigmabar (\Reff) = \frac{\Mbar(r<\Reff)}{2\pi \Reff^2}.
    \label{eq:sigmabar}
\end{equation}

Our results are shown in Fig. \ref{fig:fdm-sigmabar}, colour-coded by the stellar mass and the gas fraction. The baryonic surface densities of our galaxies span a narrow range of $\logsigmabar \approx 8-9$. We find a strong anti-correlation between $\fDM$ and $\log \sigmabar$, with Spearman rank correlation coefficient of $\rho = -0.557$ and p-value of $p<0.001$. 

We fit our $\fDM-\log\sigmabar$ relation with a power law, replicating the one used for the $\fDM-\logMstar$ (Eq. \ref{eq:power-law}). This parametrization is motivated by studies of this relation at lower redshift \citep{Wuyts:2016aa, Genzel:2017aa, Nestor-Shachar:2023aa}, to provide a more straightforward comparison. We present our best-fit parameters in Tab. \ref{tab:best-fit-sigmabar} and plot our best-fit relation in Fig. \ref{fig:fdm-sigmabar}. It is clear that our sample does not follow a clear power-law shape, which is also highlighted by the large intrinsic scatter $\sigma_{\rm int} = 0.12\pm 0.03$. The fit is also driven by the higher $\fDM$, which have smaller uncertainties. This causes most of our lower $\fDM$ values to have smaller $\sigmabar$ than predicted by the best-fit curve. 

\begin{table}
\centering
\begin{tabular}{c|c|c}
Fit & This work & This work + RC100\\
\hline
$\alpha$ & $0.48\pm 0.08$ & $0.28\pm 0.02$ \\
\hline 
$\beta$ & $-0.28\pm 0.04$ & $-0.33\pm 0.02$ \\
\hline 
$\gamma$ & $-0.05\pm 0.58$ & $-0.38\pm 0.07$ \\
\hline
$\sigma_{\rm int}$ & $0.12\pm 0.03$ &$0.11 \pm 0.01$\\
\end{tabular}
\caption{Summary of the parameters for the power-law fit to the $f_{\text{DM}}-$\sigmabar\ relation (where the parameters are defined as in Eq. \ref{eq:power-law}), as shown on Fig. \ref{fig:fdm-sigmabar}.}
\label{tab:best-fit-sigmabar}
\end{table}

The tight relation between DM fraction and $\sigmabar$ has been found to hold out to cosmic noon in observations \citep{Wuyts:2016aa, Genzel:2020aa, Nestor-Shachar:2023aa} as well as in simulations \citep{Ubler:2021aa}. This trend links the baryons with underlying the DM, and could be explained by a few different phenomena. The high baryon densities could drive out the DM in the central region through heating. Also, the high densities could be induced by a change in size of the galaxy. In general, for a fixed DM halo and at fixed stellar mass, a higher density of baryons in the centre will naturally result in a lower DM fraction when compared to the same galaxy but more diffuse.  

Our measured correlation ($\rho \approx 0.56$) is weaker than the one reported by \citet{Nestor-Shachar:2023aa} at $z\sim 2$ ($\rho = 0.64$) and $z\sim 1$ ($\rho = 0.84$). The weakening of the correlation with redshift, as was already observed at cosmic noon, points to an increase of galaxy-to-galaxy diversity. In Fig. \ref{fig:fdm-sigmabar}, we plot the best-fit relation from \citet{Wuyts:2016aa}, and we find that our galaxies with the highest DM fractions lie above this relation, as also highlighted by our best-fit relation. At lower DM fractions, our sample aligns better with the relation. 

We also perform a fit combining our $z=4-6$ data with the $z=1-2.5$ RC100 data from \citet{Nestor-Shachar:2023aa} to obtain better constraints for the redshift evolution of the $\fDM-\sigmabar$ relation. Our best-fit parameters are shown on Tab. \ref{tab:best-fit-sigmabar}. In our initial fit, the redshift dependence was poorly constrained ($\gamma = -0.05 \pm 0.58$), but when we include the RC100 data, we can constrain the increase of $\fDM$ with redshift ($\gamma = -0.38 \pm 0.07$). We investigate the origin of this increase of $\fDM$ at fixed $\sigmabar$ through the colour-coding by $\Mstar$ and $\fgas$ in the two panels of Fig. \ref{fig:fdm-sigmabar}, and comparing our sample to the \citet{Nestor-Shachar:2023aa} sample at cosmic noon. We do not colour-code by $\Reff$ because there is little overlap in sizes between the two samples, with our galaxies spanning $\Reff \approx 0.3 - 3$ kpc and the RC100 sample $\Reff \approx 3 - 10$. The RC100 sample has masses $\logMstar \approx 10-11$, gas fractions typically $\fgas\lessapprox 0.8$, and $\Reff > 3$ kpc. It is clear that our sample occupies a different parameter space, with lower masses, higher gas fractions, and smaller sizes. Interestingly, our baryonic surface densities $\log \sigmabar \approx 8-9$ have significant overlap with those of cosmic noon galaxies. This implies high densities $\sigmabar$ at high redshift, in part driven by the smaller sizes $\Reff$ \citep[e.g.][]{Shibuya:2015aa, Ward:2024aa}. In addition, the gas content is higher and more centrally concentrated, further increasing the baryonic densities in the inner regions. The shift to higher $\fDM$ at fixed $\sigmabar$ could therefore be driven by the smaller stellar masses, a direct consequence of the stellar-to-halo-mass relation where $\Mstar/M_{\rm halo}$ decreases with decreasing $\Mstar$ \citep{Moster:2010aa, Behroozi:2010aa,Tacchella:2018aa, Behroozi:2019aa}.

The high baryonic surface densities within $\Reff$ at high redshift are consistent with the build-up of bulges. In \citet{Danhaive:2025ab}, the study of multi-wavelength sizes and their ratios at $z\sim 5$ supports the growth of galaxies through central starbursts instead of smooth inside-out growth. This is consistent with bulges forming in these relatively low-mass galaxies. At cosmic noon, star-forming galaxies have been shown to predominantly grown inside-out \citep{Nelson:2016wo}, consistent with stars forming in gas disks around an older stellar bulge. The formation of these stellar disks is also reflected in the rapid increase of sizes ($\Reff$) at cosmic noon \citep{Shibuya:2015aa}. In this context, the decrease of $\fDM$, at fixed $\sigmabar$, with cosmic time reflects the mass build-up of bulges which begin to dominate the central regions of the galaxy. Also, the baryons extend further out into the halo, through inside-out growth, where the DM density decreases, which further decreases $\fDM$. 

Assuming galaxies grow along the SFMS, a $\logMstar = 9$ galaxy from our sample at $z\sim5$ would grow to $\logMstar\approx 10.5-11$ by cosmic noon, similar to the RC100 galaxies. If they maintain high DM fractions in their cores, they would evolve into a separate population from these cosmic noon galaxies, more akin to late-type galaxies (LTGs) in the local Universe \citep{Barnabe:2012aa,Martinsson:2013aa}. The subset of the galaxies in our sample with $\fDM<0.5$ would instead be candidate progenitors for the baryon-dominated galaxies at cosmic noon, having also similar gas fractions, as they occupy a similar region on the $\fDM-\sigmabar$ plane. As discussed in \citet{Nestor-Shachar:2023aa}, these galaxies are expected to become ETGs seen in the local Universe \citep{Cappellari:2013aa}. These different evolutionary tracks are consistent with the large scatter seen in our sample, suggesting a variety of different populations. However, DM fractions are not expected to stay constant within a galaxy's evolutionary track. In fact, their dependence on mass (Fig. \ref{fig:fractions-mass}) implies that as galaxies grow, their DM fraction changes. Also, the effective radii increase with cosmic time, which also affects the measurements. We investigate the redshift evolution of $\fDM$ in the next section.

\subsection{Redshift evolution of $f_{\rm DM}$} \label{sec:z-evol}

\begin{figure}
    \centering
    \includegraphics[width=1\linewidth]{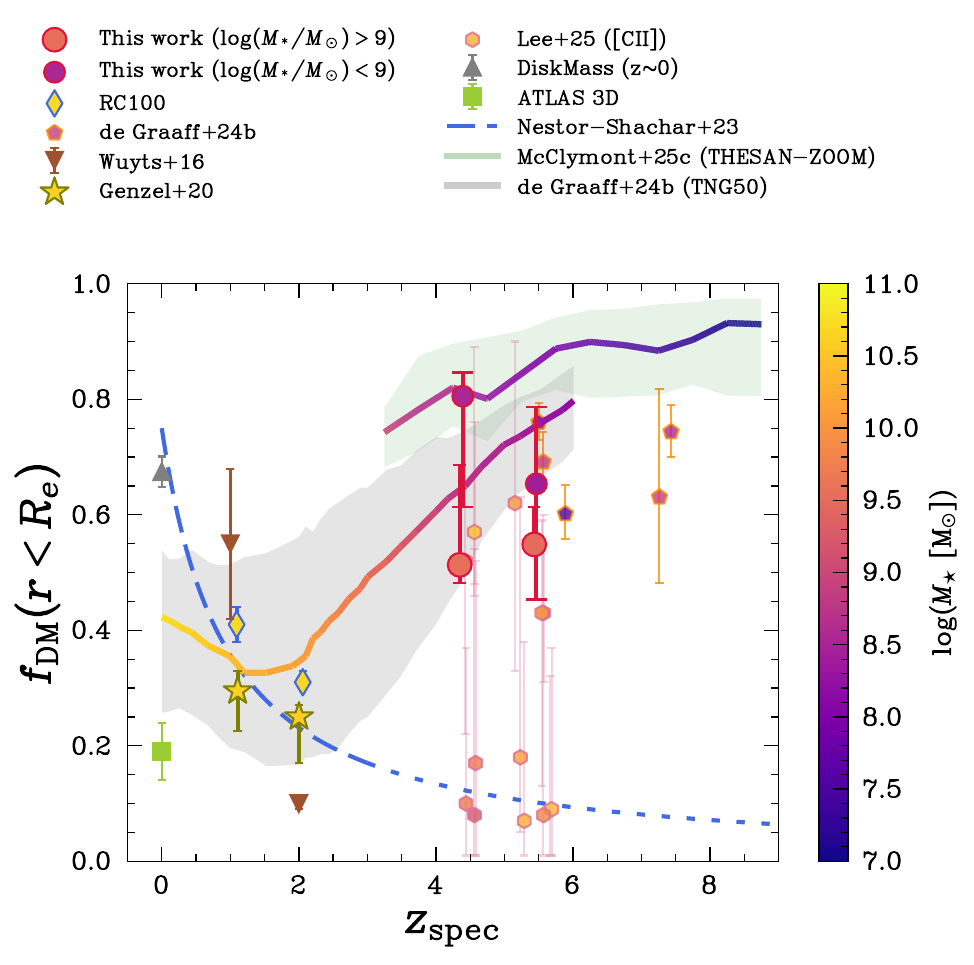}
    \caption{DM fraction as a function of redshift for our sample (circles) and samples from the literature \citep{Nestor-Shachar:2023aa,de-Graaff:2024ab,de-Graaff:2024aa}. We plot the evolution measured in \citet{Nestor-Shachar:2023aa} (dashed blue line), and its extrapolation to $z>2.5$ (dotted blue line). Our sample medians lie well above those at cosmic noon, although our sample shows large diversity in $\fDM$ and probes $\sim 2-3$ dex lower stellar masses. Qualitatively, we compare our medians ($\logMstar\approx 9$) with the median evolution of $\fDM(r<\Reff)$ for \textsc{tng50} galaxies \citep{de-Graaff:2024aa} selected at $z=6$ with $8<\logMstar<9$ (thick line with $16^{\rm th}$ and $84^{\rm th}$ percentiles shown in the grey shaded region). Similarly, we plot the median evolution for \textsc{\textsc{thesan-zoom}} galaxies \citep{McClymont:2025ac} with $\logMstar = 7 \pm 0.5$ at $z=9$ (thick line with $16^{\rm th}$ and $84^{\rm th}$ percentiles shown in the green shaded region). Although we expect redshift evolution of $\fDM(<\Reff)$ due to the evolution of sizes with redshift, the strong dependence of stellar mass is evident here.}
    \label{fig:fdm-redshift}
\end{figure}

In Fig. \ref{fig:fdm-redshift}, we plot our inferred DM fractions as a function of redshift. We plot our sample medians at $z = 4.5$ and $z=5.5$, for the lower-mass ($\logMstar<9$) and higher-mass ($\logMstar>9$) galaxies. Although our sample spans a wide range of $\fDM$, we find high fractions on average ($\fDM>0.5$). For the high-mass galaxies, with median $\logMstar=9.5$, we find $\fDM = 0.51^{+0.17}_{-0.03}$ and $\fDM = 0.55^{+0.06}_{-0.01}$ at $z = 4.5$ and $z=5.5$, respectively. For the low-mass galaxies, with median $\logMstar=8.5$, we find $\fDM = 0.81^{+0.04}_{-0.19}$ and $\fDM = 0.65^{+0.13}_{-0.20}$. In both cases, we do not see evidence for a redshift evolution, as our medians at $z = 4.5$ and $z=5.5$ are consistent within the uncertainties. This is expected given our poorly constrained $\gamma$ values (Tab. \ref{tab:best-fit-fracs}). 

We now place our measurements in the context of work at cosmic noon and the local Universe.  In the same redshift range $z\approx 4-6$, \citet{Lee:2025aa} report a wide range of $\fDM$ for their higher mass sample ($\logMstar\approx 9-10.5$), but their higher $\fDM$ systems are consistent with our high-mass medians. However, they also report galaxies that have large baryon contents ($\fDM \lessapprox 0.2$). We also find similar systems in our sample (Fig. \ref{fig:fractions-mass}) even though our medians lie at higher $\fDM$. The systems from \citet{Lee:2025aa} are therefore consistent with being a subset of the more extensive population probed in this work. At higher redshift ($z\approx 5-8$), measurements from \citet{de-Graaff:2024ab,de-Graaff:2024aa}, who find $\fDM>0.5$ for their sample of $\logMstar\approx 8-9$, are broadly consistent with our low-mass medians. On the other hand, the medians for $\logMstar\approx 10-11$ galaxies at cosmic noon \citep{Genzel:2020aa, Nestor-Shachar:2023aa} show low DM fractions ($\fDM \approx 0.2-0.4$). When combining all of these measurements together, the resulting picture suggests that mass plays a significant role in determining the DM content within the central regions of galaxies. The massive systems at cosmic noon are predominantly baryon-dominated, as is also seen in more massive systems at $z\sim 5$. On the other hand, low-mass galaxies seem more DM-dominated.

To further investigate this, we plot the median tracks from the \textsc{tng50} \citep{de-Graaff:2024aa} and \textsc{thesan-zoom} \citep{McClymont:2025ac} simulations. Specifically, for \textsc{tng50}, we plot the median evolution of $\fDM(r<\Reff)$ for galaxies selected at $z=6$ with $8<\logMstar<9$. The \textsc{thesan-zoom} track is computed as a median from galaxies with $\logMstar = 7 \pm 0.5$ at $z=9$. This comparison with simulations is qualitative, as the radii $\Reff$ are defined and measured differently in observations and simulations. Observationally, we are only able to measure the 2D half-light radius, whereas simulations define $\Reff$ as the 3D stellar half-mass radius. The choice of $\Reff$ will have an impact on the measured fractions and the derived comparisons. Despite this not being an apples-to-apples comparison, it is informative to explore its implications.

As discussed in \citet{de-Graaff:2024aa}, \textsc{tng50} predicts that low-mass galaxies are DM dominated at every epoch, and the observed evolution of $\fDM$ is driven by the masses probed. In fact, the tracks on Fig. \ref{fig:fdm-redshift} show that a single galaxy goes from being DM-dominated at $z\sim 6$ to being baryon-dominated at $z\sim 2$. This is also consistent with the tracks from \textsc{thesan-zoom}, although these simulations only reach $z=3$. This has strong implications for the evolutionary tracks of galaxies, since it suggests that the DM-dominated low-mass galaxies probed at high redshift are direct progenitors of the baryon-dominated cosmic noon galaxies. From cosmic noon, the DM-dominated systems settle as the DM-dominated LTGs in the local Universe, whereas the baryon-dominated systems could deplete their gas and settle in to the ETGs \citep{Nestor-Shachar:2023aa}. A detailed forward-modelling of the simulations is needed to conduct a better comparison with the increasing number of observed fractions.

\section{Discussion} \label{sec:discussion}

The study of DM fractions, within the effective radius, at cosmic noon ($z\approx1-3$) unveiled that the massive ($\logMstar = 10-11$) star-forming galaxy population is predominantly baryon-dominated \citep{Genzel:2017aa, Genzel:2020aa,Nestor-Shachar:2023aa}. In this work, we push the study of $\fDM$ to $z\approx 4-6$, where we are directly probing progenitors of these cosmic noon systems. In Sec. \ref{sec:fracs-mass} and Sec. \ref{sec:sigmabar}, we presented our DM fractions and their dependence on stellar mass and baryonic surface density, investigating the physical drivers for the offsets we observe with respect to cosmic noon galaxies. In Sec. \ref{sec:z-evol}, we put our measured $\fDM$ in the context of galaxies across cosmic time and predictions from tracks of cosmological simulations. 

In this section, we qualitatively explore what our measurements can teach us about the underlying DM halo, and specifically what constraints we can place on the shape of its density distribution in the central regions (Sec. \ref{sec:alpha}). From this, we hope to shed light on the consistency of different estimations of the galaxy-halo connection. We also discuss the different sources of uncertainties in our baryonic (Sec. \ref{sec:caveat-mass}) and DM (Sec. \ref{sec:caveat-DM}) mass measurements. Finally, we also consider the contribution of black holes to our measured fractions in Sec. \ref{sec:mbh}.

\subsection{Empirical model predictions} \label{sec:alpha}
\begin{figure*}
    \centering
    \includegraphics[width=1\linewidth]{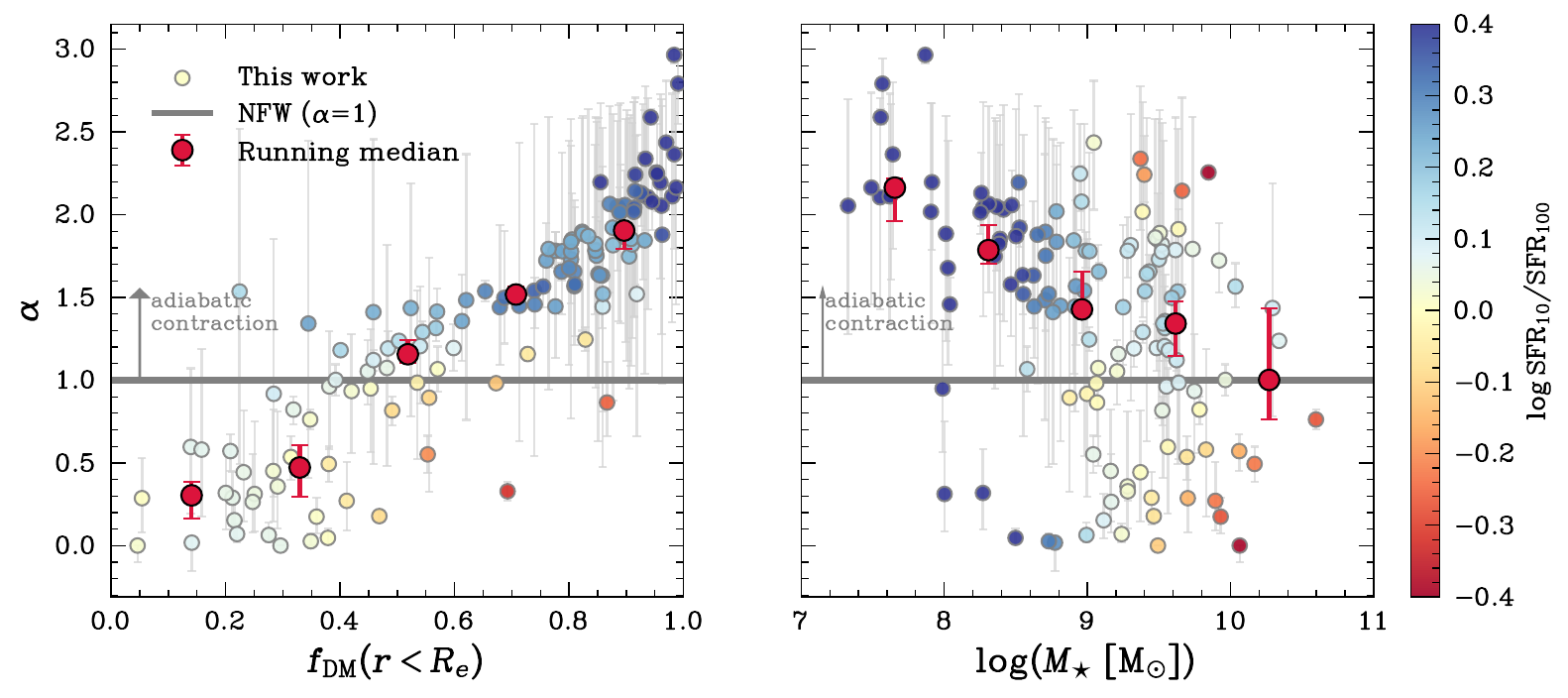}
    \caption{Estimated slope $\alpha$ of the inner DM density profile for a modified NFW distribution (Eq. \ref{eq:gen-nfw}) as a function of measured DM fractions (left) and stellar mass (right). A value of $\alpha = 1$ represents the classical NFW profile, whereas $\alpha < 1$ corresponds to a cored profile, and $\alpha>1$ to a cuspier profile. Under the assumed SHMR and concentration ($c=5$), we find that low-mass galaxies with high $\fDM$ are associated with cuspy distributions, converging towards an NFW-like profile at higher masses with a scatter at least in part driven by their star-formation state. To highlight this, we colour-code the galaxies by the ratio of their SFR averaged over 10 and 100 Myr, $\rm SFR_{10}/SFR_{100}$, applying \textsc{LOESS} smoothing \citep{Cappellari:2013ac}. Galaxies going through a burst of star formation ($\log \rm SFR_{10}/SFR_{100} \gtrapprox 0.4$) have preferentially high values of $\alpha$, which supports the adiabatic contraction scenario.}
    \label{fig:alpha_fdm_sfr}
\end{figure*}

Our modelling, which is limited to probing rotational curves out to $\approx 1-2 ~\Reff$, does not allow for a detailed recovery of the underlying DM density profiles of our galaxies. Ideally, rotation should be probed further out to several $\Reff$. However, the measure of the DM fractions within $\Reff$ can provide constraints on the DM density profile via comparisons to empirical predictions, which require various assumptions. We explore these constraints in this section, but caution that they remain qualitative and reliant on assumed relations.

The stellar-to-halo mass (SMHM) relation has been studied in detail both in empirical models \citep{Behroozi:2019aa, Tacchella:2018aa} and in simulations \citep{Ceverino:2017aa,Ma:2018aa}, with constraints from observations \citep{Kravtsov:2018aa, Girelli:2020aa, Shuntov:2022aa, Shuntov:2025aa}. This relation links the halo mass of galaxies to their stellar mass at a given redshift, allowing us to estimate the halo masses for the galaxies in our sample. In order to estimate the halo masses for our galaxies, we adopt the \citet{Behroozi:2019aa} SMHM relation, which has a double power-law shape with an added Gaussian:

\begin{equation}
    \log_{10}\left(\frac{M_*}{M_1}\right) = \varepsilon - \log_{10}\left(10^{-\alpha x} + 10^{-\beta x}\right) + \gamma \exp\left[-0.5\left(\frac{x}{\delta}\right)^2\right]
    \label{eq:smhm}
\end{equation}
where $x \equiv \log_{10}\left(\frac{M_{\text{peak}}}{M_1}\right)$, $M_{\text{peak}}$ is the peak halo mass and $M_1$ is a characteristic mass. The free parameters $M_1$, $\epsilon, \alpha, \beta$ and $\gamma$ all have redshift-dependent expressions, and $\delta$ is a constant. Once we have inferred the halo mass by numerically solving Eq. \ref{eq:smhm}, we can derive the corresponding virial radius:
\begin{equation}
    R_{\rm vir} = \left( \frac{4}{3} \frac{\pi 200\rho_c}{M_{\rm halo}}\right)^{-1/3},
\end{equation}
since $R_{\rm vir}$ is defined as the radius where the density is 200 times the critical density $\rho_c$ of the Universe at that redshift, and $M_{\rm halo}$ is the DM mass within the virial radius. For the shape of the DM density profile, we assume a generalized Navarro–Frenk–White \citep[NFW,][]{Navarro:1997aa} profile: 

\begin{equation}
    \rho_{\text{DM}}(r) = \frac{\rho_0}{(r/r_s)^{\alpha}(1 + r/r_s)^{3-\alpha}},
    \label{eq:gen-nfw}
\end{equation}
where $\alpha = 1$ corresponds to a classic NFW profile. Given these assumptions, we can constrain the value of $\alpha$ needed to reproduce our measured DM fractions $\fDM$ within the effective radius. We restrict our values between $\alpha = 0$ and $\alpha =3$, above (below) which the profile becomes convex (concave) in logarithmic space. In fact, for $\alpha > 3$, Eq.~\ref{eq:gen-nfw} develops an upturn near $r = r_s$, as the outer slope $(3 - \alpha)$ becomes negative. Conversely, when $\alpha < 0$, the inner slope ($\alpha$) becomes negative, producing a downturn at $r = r_s$. Within the allowed range $0 \leq \alpha \leq 3$, the profile smoothly transitions from a nearly flat core ($\alpha = 0$) to a progressively steeper slope that becomes (log-)linear at $\alpha = 3$. 

The only additional unknown is the concentration $c$, defined as
\begin{equation}
    R_{\rm vir} = cR_{\rm s}
\end{equation}
where $R_{\rm s}$ is the scale radius. Given our other assumptions, the value of $c$ effectively re-normalizes the predicted DM fractions for all of our systems, increasing them as $c$ increases. DM only simulations predict low concentrations at $z\sim 5$ \citep{Dutton:2014aa}, so we set $c=5$. 

In Fig. \ref{fig:alpha_fdm_sfr}, we present the estimated values of $\alpha$ needed to reproduce our measured DM fractions within the effective radius. We find that $\approx 30\%$ of our sample lies below $\alpha = 1$, implying that their DM fractions are best reproduced by a profile that is more cored than the classical NFW. Evidence for DM cores has already been found at cosmic noon \citep{Genzel:2020aa, Nestor-Shachar:2023aa}, where samples are predominantly baryon-dominated and require cored profiles to reproduce their low $\fDM$. Aside from more extreme solutions such as the modification of the nature of DM \citep{Hu:2000aa, Calabrese:2016aa} or of our gravity model \citep{Milgrom:1983aa}, a more natural explanation for such cores is a strong interaction between the baryons and the DM within galaxies. Specifically, this would require kinetic heating of the DM in the central regions to drive it outwards. This could be achieved through effective feedback mechanisms from star formation and black holes. Another possibility is that some of these low $\fDM$ galaxies are in a post-merger phase, with disturbed kinematics and diffused DM haloes. This is plausible given that most of the galaxies in the $\alpha<0$ region have higher stellar masses ($\logMstar>9$). 

On the other hand, high central baryonic concentrations are expected to perturb the underlying DM distribution and pull the DM towards the centre, increasing the "cuspiness" of its density profile. This phenomenon is called adiabatic contraction \citep{Blumenthal:1986aa}, and is represented by the $\alpha>1$ region of Fig. \ref{fig:alpha_fdm_sfr}. For galaxies in this region, adiabatic contraction is needed to explain the high $\fDM$ values measured in this work. Interestingly, this under-prediction of $\fDM$, compared to our observations, holds for many other SMHM relations from the literature which typically lie above the \citet{Behroozi:2019aa} one \citep[e.g.][]{Tacchella:2018aa}. In order to bridge some of the discrepancies between the model predictions and our measurements, a lower SMHM relation at high redshift would be needed. This problem is emphasized by recent predictions from observations and simulations of high DM fractions in low-mass galaxies \citep{de-Graaff:2024aa, McClymont:2025ac}. However, it is important to note that our high $\alpha\gtrapprox1.5$ values suffer from large uncertainties that make them consistent with $\alpha \approx 1.5$. These stem from the smaller masses of these systems, which make their morphology and kinematics more difficult to constrain.

We colour-code our points in Fig. \ref{fig:alpha_fdm_sfr} by their SFHs, parametrized by the ratio of SFR averaged over $10$ and $100$ Myr. Galaxies with rising SFHs will have positive values of $\rm SFR_{10}/SFR_{100}$, whereas galaxies with falling SFHs will show the opposite behaviour. In order to visually asses the presence of an underlying trend, we combine our points in larger bins and use the LOESS method \citep{Cappellari:2013ac} to average over single objects and obtain mean estimates for the full sample. We can see a trend of galaxies going through a burst $\log \rm SFR_{10}/SFR_{100} \gtrapprox 0.4$ having preferentially higher values of $\alpha$. This apparent correlation supports the adiabatic contraction scenario that would occur with the inflow of baryons to the central regions during a burst, during which the gas content of the galaxy contracts \citep{Dekel:2014aa,Tacchella:2016aa,McClymont:2025ab}. When feedback from the star formation kicks in, driving the gas out of the centre and temporarily quenching star formation, galaxies begin to move back down towards the MS $\log \rm SFR_{10}/SFR_{100} <0$. These galaxies appear to preferentially have cored profiles, consistent with kinetic heating of the DM. This interpretation would imply the interaction of baryons and DM on relatively short timescales ($t\approx 100$ Myr), which has yet to be thoroughly explored in cosmological simulations. Interestingly, we also find that the values of $\alpha$ correlate with the \sersic index $n$ measured from the stellar light distribution in the UV (Fig. \ref{fig:alpha_fdm_mstar_n}). Galaxies with cuspier DM profiles also have steeper central light profiles, with $n\sim4$, which could point to interactions between the two components.

We find that low-mass galaxies $\logMstar<9$ have cuspier profiles, while higher mass galaxies seem to converge towards NFW profiles ($\alpha = 1$) with scatter driven by their star-formation state (parametrized by $\rm SFR_{10}/SFR_{100}$). This is not necessarily intuitive given the initial discovery of cored systems in dwarf galaxies in the local Universe. However, this can be reconciled when considering burstiness. At high redshift, galaxies have been shown to grow through bursts of star formation \citep[e.g.][]{Faucher-Giguere:2018aa,Tacchella:2020aa,McClymont:2025aa, Simmonds:2025aa} at all masses. These bursts are tied to the formation of cores through energetic stellar feedback smoothing the DM profile in the central regions of galaxies. The more massive galaxies in our sample will have undergone more bursts in their lifetime, leading to a flattening of their DM profiles and resulting in the more cored profiles we predict in this work. Also, these galaxies could undergo breathing cycles of compaction phases \citep{Tacchella:2016ab, El-Badry:2016aa}. Interestingly, \citet{Kohandel:2025aa} find that one of their most massive galaxies ("Amaryllis", $\logMstar=10.3$ at $z=7$) has a cored inner DM density profile which reflects the cusp–flattening impact of baryonic feedback.

In the local Universe, star formation is smoother, with massive star-forming galaxies growing more smoothly inside out once their central bulges have developed. These massive galaxies undergo long-lived stable growth episodes that could allow their DM profile to stabilise around NFW profiles. On the other hand, dwarf galaxies still have bursty star formation \citep{Hopkins:2014aa, Hayward:2017aa, Hopkins:2023aa}, giving rise to their cores. It is also important to once again consider the evolution of sizes. These dwarf galaxies have larger radii than low-mass galaxies at the same mass in our sample. This naturally implies higher concentrations, which could fuel processes of adiabatic contraction for which we see evidence in Fig. \ref{fig:alpha_fdm_sfr}.

Finally, we assess how sensitive our results are to the adopted assumptions and discuss their potential impact on our conclusions. Starting with the concentration, changing the value of $c$ affects the distribution of $\alpha$ we find. Specifically, higher concentrations $c>5$ require more galaxies to have cored profiles $\alpha<1$ in order to reconcile the predicted fractions $\fDM$ with our measured ones. In contrast, lower concentrations, $c<3$, would have the opposite effect. The most important choice is the SMHM function, which directly affects the normalization of our profiles and, hence, of the $\alpha$ values that we infer. Studies based on simulations and empirical models at high redshift suggest that the normalization of the SMHM relation is higher than the \citet{Behroozi:2019aa} one assumed in this work \citep[e.g. see][]{Tacchella:2018aa}. This means that the halo mass $M_{\rm halo}$ at fixed stellar mass $\Mstar$ is lower than predicted by the \citet{Behroozi:2019aa} relation, meaning our profiles would need to be even cuspier (higher $\alpha$) to reproduce our high $\fDM$ values. A last assumption is the choice of profile for the fit, as other profiles such as the Einasto profile \citep{Einasto:1965aa}, have also been shown to provide good fits to DM profiles in cosmological simulations \citep{Dutton:2014aa}. We explored this profile and found no significant change in the observed trends of cuspiness with $\fDM$, $\rm SFR_{10}/SFR_{100}$, and $\Mstar$. However, our points become more scattered in these parameter spaces. We chose to study the modified NFW profile in the most detail because of the direct comparison it offers with the NFW ($\alpha = 1$) case (Eq. \ref{eq:gen-nfw}).

Although it is informative to discuss empirical model predictions in the context of our results, it is also important to discuss caveats of the measurements that could potentially reconcile our inferred profiles with shapes closer to NFW. We discuss these in the next two sections, focusing on the uncertainties in the baryonic and the DM contents.

\subsection{Uncertainties in the baryonic content}\label{sec:caveat-mass}

When calculating baryonic masses needed to infer gas and DM fractions, we need to make some assumptions, of which we will now discuss the implications. First, the stellar mass is inferred using the SED-fitting code \textsc{Prospector}, a method with many advantages, but also some degeneracies. We attempt to break some of the degeneracies by fixing the redshift to the grism spectroscopic redshift, and by simultaneously fitting the photometry with the line fluxes of the available emission lines (see Sec. \ref{sec:methods}).  Despite this, we are not fitting spectra and do not have strong constraints for all emission lines, for instance, those constraining nebular metallicities and dust content. Furthermore, our S/N cut implies selecting galaxies whose young stellar population is dominating the SED. This may hamper the detection of the underlying population of older stars, and hence constraining the full SFH. This effect is called 'outshining' and can cause an underestimate of the stellar masses, effectively only attributing the stellar mass to the more visible young stellar population \citep{Bell:2001aa,Maraston:2010aa,Leja:2019ab,Tacchella:2023aa,Gimenez-Arteaga:2023aa}. This would directly affect the measurements of $\fDM<0$, alleviating the tension by increasing the total baryonic mass.

Uncertainties in the stellar mass and SFRs are propagated into the gas mass estimates derived using the scaling relation by \cite{Tacconi:2018aa, Tacconi:2020aa}. This relation is calibrated out to $z\sim5.5$, and should hence provide a better estimate of the gas in our galaxies than relations calibrated at much lower redshifts  \citep[e.g.][]{Kennicutt:1998vu}. Nonetheless, the use of scaling relations does not necessarily encompass the large variety of gas fractions observed at fixed mass \citep[e.g.][]{McClymont:2025ac}, and could introduce biases. Importantly, the \citet{Tacconi:2020aa} relation is only calibrated above $\logMstar\gtrapprox 9$. The extrapolation to the lower masses in our sample results in very high gas fractions $\fgas\sim 1$.

\subsection{Uncertainties in the DM content} \label{sec:caveat-DM}

The kinematic measurements, described in \citet{Danhaive:2025aa}, are the main factor driving the measurement of $\fDM$ through their estimate of the dynamical mass. The key assumption made when deriving the dynamical mass is that the measured velocity gradients are tracing rotation of the gas around the galaxy. However, especially at low masses, we cannot rule out the contribution of non-circular motions, typically in the form of outflows \citep{Carniani:2024aa,Ivey:2025aa}, to our measurement of velocity gradients and dispersions. Such a contribution would bias both quantities to higher values and unphysically boost the inferred dynamical masses, and hence the DM fractions. Also, albeit of smaller importance, the circular velocities needed to compute the dynamical masses are inferred assuming a virialised rotating disk with an exponential light profile. For galaxies with larger S\'{e}rsic indices, this assumption can lead to biased results. Furthermore, for pressure dominated systems, the choice of the pressure support term multiplying the $\disp$\ in Eq. \ref{eq:v_circ} can lead to over or under-estimates  of the circular velocity \citep[see][for detailed analysis]{Price:2022aa}. 

For the derived relations for both the gas and importantly the DM fractions, we note that we suffer from incompleteness at the low-mass end ($\logMstar<9$), which could significantly bias our results. In order to obtain meaningful (i.e. resolved) kinematic measurements for low-mass systems, they need to be not only bright in \Ha, but also have relatively large rotational velocity and/or velocity dispersions. This could cause our $\fgas$ to be biased high (due to the high SFR requirement for the bright \Ha) as well as our $\fDM$ (due to the high circular velocities). Accounting for these selection effects would move our observed medians closer to the simulations in Fig. \ref{fig:fractions-mass}. Within the paper, we discuss the observed trends, but cannot make definitive conclusions about the behaviour at the low masses. This also translates to incompleteness for galaxies with low baryonic surface densities. 

In this discussion, there is a last component that we have not considered, namely the presence of BHs in our galaxies. We investigate this possibility in the next section.

\subsection{Overmassive BHs} \label{sec:mbh}

\begin{figure}
    \centering
    \includegraphics[width=1\linewidth]{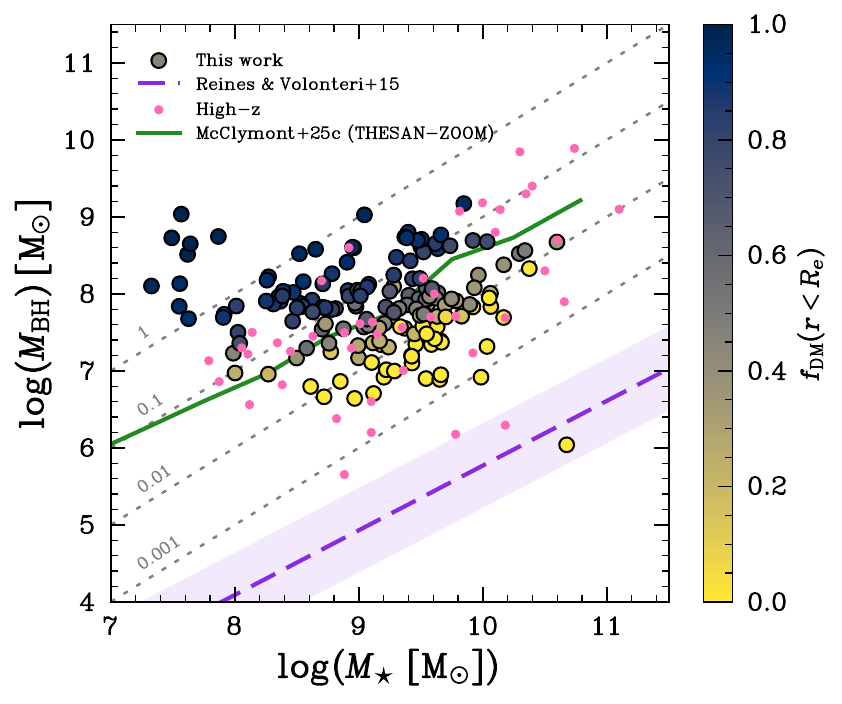}
    \caption{Predicted black hole masses $M_{\rm BH}$, assuming the local $M_{\rm BH}-M_{\rm dyn}$ relation from \citet{Kormendy:2013aa}, as a function of stellar mass (circles). We compare our measurements with the overmassive black holes reported at high redshift ($z\approx 4-11$) from \textit{JWST} \citep[pink dots;][]{Carnall:2023aa, Kokorev:2023aa,Harikane:2023aa, Matthee:2024aa,Ubler:2023tn, Maiolino:2024aa,Maiolino:2024ab,Ubler:2024aa, Greene:2024aa,Furtak:2024aa,Juodzbalis:2024aa,Natarajan:2024aa}, which lie well above the local relation \citep[purple dashed line;][]{Reines:2015aa}. The dashed grey lines represent constant ratios of $\Mbh/\Mstar$. We find excellent agreement with the parameter space spanned by these sources, suggesting the shift in the $M_{\rm BH}-\Mstar$ relation is consistent with the high gas and DM fractions found in high redshift galaxies. This is consistent with results from \citet{McClymont:2025ac} following a similar approach to this work but with the \textsc{thesan-zoom} simulations.}
    \label{fig:mbh-mstar}
\end{figure}

 Although BHs typically represent a small fraction of the total mass, with $M_{\rm BH}/\Mstar<1\%$ at $z\sim 0$ \citep{Reines:2015aa}, they could have non-negligible effects on the kinematics of their host galaxies through potentially disruptive feedback mechanisms. Furthermore, many recent studies have reported over-massive BHs in the early Universe, with high $M_{\rm BH}/\Mstar$ ratios \citep{Carnall:2023aa, Kokorev:2023aa,Harikane:2023aa, Ubler:2023tn, Maiolino:2024aa,Maiolino:2024ab,Ubler:2024aa, Furtak:2024aa,Natarajan:2024aa,Juodzbalis:2025ab} and in some cases even high $M_{\rm BH}/\Mdyn$ ratios \citep{DEugenio:2025aa,Ji:2025aa,Juodzbalis:2025aa} relative to the local scaling relations. 

In order to investigate the presence of BHs in our galaxies, and quantify their potential contributions to our measured DM fractions, we derive BH masses using the $M_{\rm BH}-\Mdyn$ relations from \citet{Kormendy:2013aa}, which has been shown to hold at high redshift for most cases \citep{Maiolino:2024aa,Juodzbalis:2025aa}. We show our results on Fig. \ref{fig:mbh-mstar}. 

We find BH masses spanning a wide range ($\logMbh = 6.5 - 9$) and BH-to-stellar mass ratios ($\Mbh/\Mstar = 0.001-1$). Interestingly, our sample overlaps with most of the over-massive BHs reported at high redshift. This suggests that, given a fixed $M_{\rm BH}-\Mdyn$ relation like we have assumed here, these over-massive black holes (relative to the $M_{\rm BH}-\Mstar$ plane) are a natural consequence of the high gas and DM fractions found at high redshift (Fig. \ref{sec:fracs-mass}). This is consistent with the results from \citet{McClymont:2025ac}, who applied a similar semi-empirical approach to the \textsc{thesan-zoom} simulations, whose model does not include black holes, to predict black hole masses in the post-processing. Our results imply that ratios of $\Mbh/\Mstar\approx 0.1$ would be common at high-redshift, and that black holes grow faster at early times, in comparison to the stellar content, than at late times. Because both growth mechanisms require gas inflow, the gas must collapse enough to accrete on to the black hole but without forming many stars. Black hole growth would need to differ at high redshift in order to reproduce these high $\Mbh/\Mstar$ ratios. However, confirming the presence of black holes at high redshift has proven difficult with our current emission line diagnostics \citep[e.g.][]{Mazzolari:2024aa, Scholtz:2025ac,Ivey:2025aa}, and various emission line ratios need to be detected spectroscopically, which we do not probe with the grism data used in this work. Also, many models expect that the bulk of the BH population at high-$z$ should be dormant \citep[e.g.][]{Schneider:2023aa,Trinca:2024aa}, with short duty cycles for accretion, meaning that we would not expect to see evidence of accreting BHs.

We note that the lower mass systems $\logMstar<8$ in our sample have abnormally high dynamical masses, leading to high predictions of $\Mbh$, pointing to a likely contribution from non-circular motions, as also shown in Sec. \ref{sec:TFR}. For the remainder of the systems, our findings imply that the BHs could have a significant impact on their host galaxies. Although including their masses in our calculation of $\fDM$ does not strongly affect these measurements, due to the small $\Mbh/\Mdyn$ values \citep{Kormendy:2013aa}, those who are accreting could affect the kinematics of the surrounding gas and introduce biases in the measurement of $\Mdyn$. If present, these AGNs could be categorized as Type-II, narrow-line AGNs, and hence would not be detected through the broadening of emission lines caused by the broad-line region. Nonetheless, AGN-driven outflows in low-mass galaxies would have similar velocities to those measured in this work \citep[$v\sim 100-200$ km/s, e.g.][]{Ivey:2025aa}, and could hence boost our measured DM fractions.

\section{Summary \& Conclusions }\label{sec:conclusions}

This work presents the dynamical mass measurements of 163 \Ha\ emitters at $z\approx 4-6$ from the FRESCO and CONGRESS \textit{JWST} grism surveys in the GOODS fields. We model the kinematics of the $\Ha$ emission line using forward-modelling and fitting of the grism data with \geko\ \citep{Danhaive:2025ac}, recovering in particular rotational velocities and velocity dispersions. Using our measured dynamical masses from our modelling, stellar masses inferred from SED modelling, and gas masses estimated from the \citet{Tacconi:2020aa} scaling relation, we obtain the gas fractions ($\fgas$) and DM fractions ($\fDM$) within the $\Ha$ half-light radius ($\Reff$). We summarize our main findings here:
\begin{itemize}

    \item The $\logMstar \approx 7-10$ star-forming galaxies in our sample have relatively high dynamical masses in the range $\logMdyn \approx 9-11$. At fixed stellar mass, there is large scatter in the dynamical masses, indicative of unsettled kinematics. This scatter naturally manifests itself in the Tully–Fisher plane, suggesting that the relation is only beginning to emerge at $z\sim5$.

    \item Our galaxies have, on average, high molecular gas and DM fractions, with sample medians $<\fgas> = 0.77$ and $<\fDM> = 0.73$. We find that $\approx 67\%$ of our galaxies are DM dominated with their $\Reff\sim 0.5-1$ kpc, $\fDM>0.5$. Nonetheless, our fractions $\fDM$ span the full range of the parameter space.

    \item We find evidence for a negative dependence of $\fDM$ on stellar mass $\Mstar$ as parametrized by Eq. \ref{eq:power-law}. Specifically, we find $\alpha = 0.28 \pm 0.05$, with low-mass ($\logMstar < 9$) galaxies showing high DM fractions and higher mass galaxies ($\logMstar > 9$) showing a larger diversity, with typically lower $\fDM$.

    \item We find an anti-correlation ($\rho = -0.56, p<0.001$) between $\fDM$ and the baryonic surface density $\sigmabar$ within $\Reff$, consistent with but weaker than its counterpart at cosmic noon and the local Universe. The galaxies in our sample have high baryonic surface densities comparable to those of more massive galaxies at cosmic noon, caused by their compactness and high central gas fractions.

    \item Our high DM fractions are consistent with the predicted progenitor populations of $z\sim 2$ baryon-dominated systems, as shown by a comparison with the \textsc{tng50} and \textsc{thesan-zoom} simulations. These high fractions are expected for low-mass galaxies at all redshifts.

    \item Assuming a modified NFW profile, a stellar-to-halo mass function, and a DM profile concentration, we qualitatively explore the predicted shape of the underlying DM halo density profile for our sample. We find that the higher-mass, baryon-dominated systems would need a cored profile to reconcile their low fractions. This core could have been induced by repeated bursts of star formation. In contrast, the low-mass, high $\fDM$ systems are consistent with cuspier DM profiles, suggesting adiabatic contraction pulling more DM into the central regions.

    \item Finally, we find that our elevated $\fgas$ and $\fDM$ naturally anticipate the population of over-massive black holes found with \textit{JWST} at high-redshift, when assuming a $\Mbh-\Mdyn$ relation. 
\end{itemize}

Our study extends, for the first time, measurements of DM fractions to large statistical samples at high redshift, advancing beyond previous analyses focused on ionised-gas kinematics at cosmic noon. Although our inferences rely on several necessary assumptions, they provide a valuable framework for placing spatially resolved measurements of individual systems into a broader population context. Most importantly, this work establishes a foundation for probing DM haloes through high-redshift observations, which is an essential step toward understanding how baryons and dark matter interact and co-evolve during the early stages of galaxy formation. 

\section*{Acknowledgements}
We thank Hannah Übler and Andreas Burkert for the insightful discussions. ALD thanks the University of Cambridge Harding Distinguished Postgraduate Scholars Programme and Technology Facilities Council (STFC) Center for Doctoral Training (CDT) in Data intensive science at the University of Cambridge (STFC grant number 2742605) for a PhD studentship. ALD and ST acknowledge support by the Royal Society Research Grant G125142. AJB acknowledges funding from the "FirstGalaxies" Advanced Grant from the European Research Council (ERC) under the European Union’s Horizon 2020 research and innovation programme (Grant agreement No. 789056). ECL acknowledges support of an STFC Webb Fellowship (ST/W001438/1). FDE and RM acknowledge support by the Science and Technology Facilities Council (STFC), by the ERC through Advanced Grant 695671 ``QUENCH'', and by the UKRI Frontier Research grant RISEandFALL. EE, BDJ, MR, CNAW, and YZ are supported by the JWST/NIRCam contract to the University of Arizona NAS5-02105. DJE is supported as a Simons Investigator and by JWST/NIRCam contract to the University of Arizona, NAS5-02105.  Support for program \#3215 was provided by NASA through a grant from the Space Telescope Science Institute, which is operated by the Association of Universities for Research in Astronomy, Inc., under NASA contract NAS 5-03127. WM thanks the Science and Technology Facilities Council (STFC) Center for Doctoral Training (CDT) in Data Intensive Science at the University of Cambridge (STFC grant number 2742968) for a PhD studentship. BER acknowledges support from the NIRCam Science Team contract to the University of Arizona, NAS5-02105, and JWST Program 3215.

This work is based on observations made with the NASA/ESA Hubble Space Telescope and NASA/ESA/CSA James Webb Space Telescope. The data were obtained from the Mikulski Archive for Space Telescopes at the Space Telescope Science Institute, which is operated by the Association of Universities for Research in Astronomy, Inc., under NASA contract NAS 5-03127 for \textit{JWST}. These observations are associated with program \#1180, 1181, 1210 (JADES), \#1895 (FRESCO), \# 1963 (JEMS) and \#3577 (CONGRESS).
Support for program \#3577 was provided by NASA through a grant from the Space Telescope Science Institute, which is operated by the Association of Universities for Research in Astronomy, Inc., under NASA contract NAS 5-03127. The authors acknowledge the FRESCO team for developing their observing program with a zero-exclusive-access period. The authors acknowledge use of the lux supercomputer at UC Santa Cruz, funded by NSF MRI grant AST 1828315.

% %%%%%%%%%%%%%%%%%%%%%%%%%%%%%%%%%%%%%%%%%%%%%%%%%%
\section*{Data Availability}

The data underlying this article will be shared on reasonable request
to the corresponding author. Fully reduced NIRCam images are publicly available on MAST (\url{https://archive.stsci.edu/hlsp/jades}), with \doi{10.17909/8tdj-8n28}, \doi{10.17909/z2gw-mk31}, and \doi{10.17909/fsc4-dt61} \citep{Rieke:2023aa, Eisenstein:2023aa}. The NIRCam grism spectra are publicly available on MAST with \doi{10.17909/6rfk-6s81} and \doi{0.17909/gdyc-7g80} \citep{Oesch:2023aa}.

% The inclusion of a Data Availability Statement is a requirement for articles published in MNRAS. Data Availability Statements provide a standardised format for readers to understand the availability of data underlying the research results described in the article. The statement may refer to original data generated in the course of the study or to third-party data analysed in the article. The statement should describe and provide means of access, where possible, by linking to the data or providing the required accession numbers for the relevant databases or DOIs.



\bibliographystyle{mnras}
\bibliography{PhD_bib} % if your bibtex file is called example.bib

%\begin{thebibliography}{99}
%\end{thebibliography}

%%%%%%%%%%%%%%%%%%%%%%%%%%%%%%%%%%%%%%%%%%%%%%%%%%

%%%%%%%%%%%%%%%%% APPENDICES %%%%%%%%%%%%%%%%%%%%%

\appendix

\section{Supplementary figures}

\begin{figure*}
    \centering
    \includegraphics[width=1\linewidth]{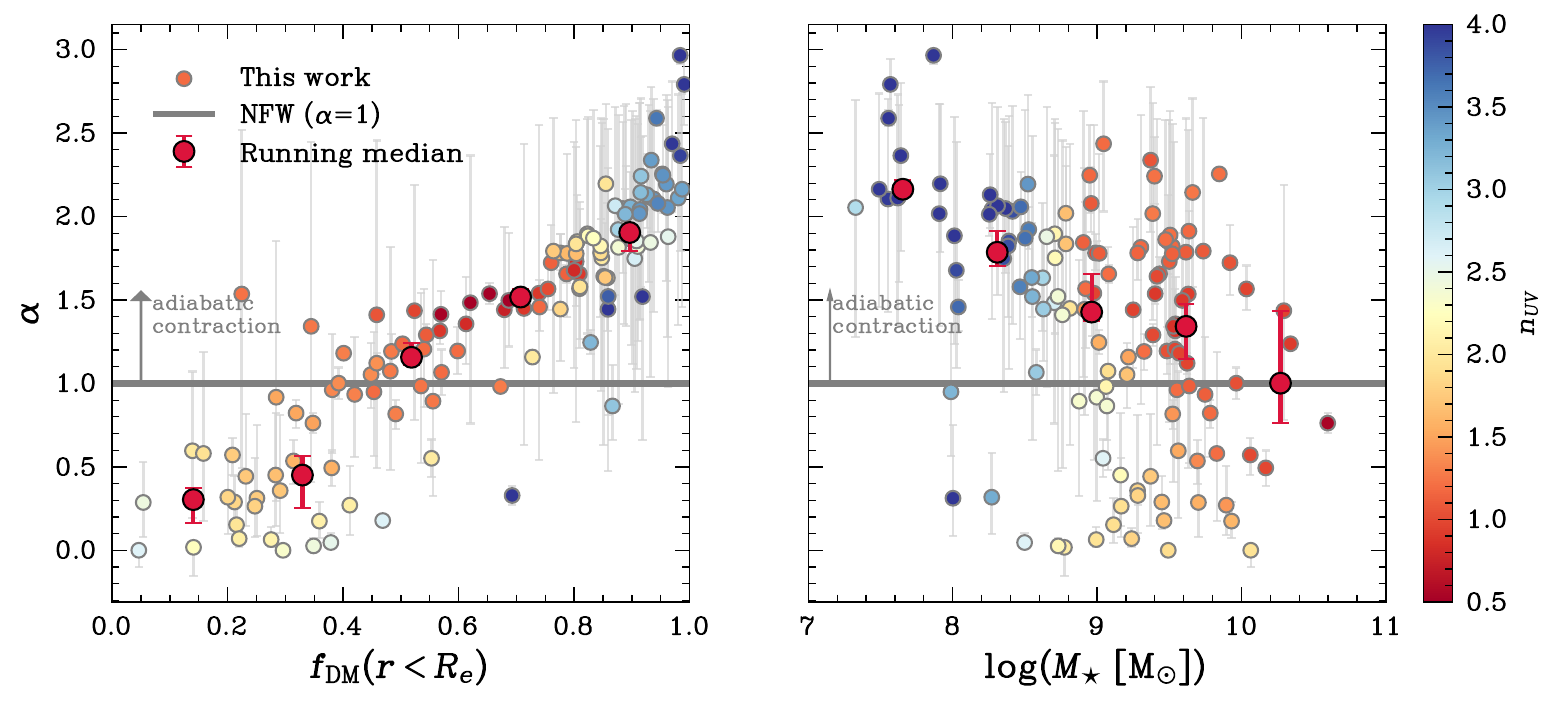}
    \caption{Estimated slope $\alpha$ of the inner DM density profile for a modified NFW distribution (Eq. \ref{eq:gen-nfw}) as a function of measured DM fractions (left) and stellar mass (right). A value of $\alpha = 1$ represents the classical NFW profile, whereas $\alpha < 1$ corresponds to a cored profile, and $\alpha>1$ to a cuspier profile. We find that galaxies with seemingly steeper DM profiles ($\alpha>1$) coincide with steeper light distributions ($n>1$) as probed by the UV stellar continuum. To highlight this, we colour-code the galaxies by the \sersic index $n_{\rm UV}$, applying \textsc{LOESS} smoothing \citep{Cappellari:2013ac}.}
    \label{fig:alpha_fdm_mstar_n}
\end{figure*}

In Figure~\ref{fig:alpha_fdm_mstar_n}, we present additional diagnostic plots that complement the main analysis. This figure illustrates the relation between the inferred inner DM slope $\alpha$ and the DM fraction $f_{\rm DM}$, colour-coded by stellar mass and Sérsic index. The trends shown here are consistent with those discussed in Section~\ref{sec:discussion}, reinforcing that galaxies with lower $f_{\rm DM}$ tend to exhibit shallower inner density slopes, while systems with higher $f_{\rm DM}$ favour cuspier profiles. Although the scatter is substantial, these correlations support the interpretation that the interplay between baryonic concentration and halo response is already shaping the inner DM structure at $z\sim5$.

%%%%%%%%%%%%%%%%%%%%%%%%%%%%%%%%%%%%%%%%%%%%%%%%%%

% Don't change these lines
\bsp	% typesetting comment
\label{lastpage}
\end{document}